\documentclass[
 reprint,
 amsmath,amssymb,
 aps,preprintnumbers,nofootinbib
]{revtex4-1}

\usepackage{graphicx}
\usepackage{dcolumn}
\usepackage{bm}
\usepackage{amsmath,amssymb,graphicx,bbm,mathrsfs,nicefrac}
\usepackage{tikz}
\usepackage{amsmath}
\usepackage{amssymb}
\usepackage{graphicx,textcomp,float,gensymb,wrapfig, enumitem,comment,dsfont,framed,slashed,appendix,wrapfig,xcolor}
\usepackage{ bbold }
\usepackage{braket} 
\usepackage{ mathrsfs }
\usepackage{caption}
\usepackage{subcaption}
\usepackage{etoolbox,hyperref,makecell}
\usepackage{feynmp}
\usetikzlibrary{arrows.meta}
\usepackage{empheq}
\usepackage{enumitem}
\usepackage[normalem]{ulem}
\setenumerate[1]{label=(\arabic*)}
\newcommand{\bea}{\begin{eqnarray}}  
\newcommand{\eea}{\end{eqnarray}}

\newcommand{\nc}{\newcommand}

\nc{\beq}{\begin{equation}}
\nc{\eeq}{\end{equation}}
\nc{\barray}{\begin{eqnarray}}
\nc{\earray}{\end{eqnarray}}
\nc{\barrayn}{\begin{eqnarray*}}
\nc{\earrayn}{\end{eqnarray*}}
\nc{\bcenter}{\begin{center}}
\nc{\ecenter}{\end{center}}
\nc{\mc}{\mathcal}
\nc{\er}[1]{(\ref{eq:#1})}
\nc{\onehalf}{\frac{1}{2}} 
\nc{\partialbar}{\bar{\partial}}
\nc{\psit}{\widetilde{\psi}}
\nc{\hc}{\mbox{H.c.}}
\nc{\ev}{\;\mathrm{eV}}
\nc{\mev}{\;\mathrm{MeV}}
\nc{\gev}{\;\mathrm{GeV}}
\nc{\kev}{\;\mathrm{keV}}
\nc{\tev}{\;\mathrm{TeV}}

\def\chii0{\chi_i^0}
\def\chij0{\chi_j^0}

\newcommand{\gsim}{\lower.7ex\hbox{$\;\stackrel{\textstyle>}{\sim}\;$}}
\newcommand{\lsim}{\lower.7ex\hbox{$\;\stackrel{\textstyle<}{\sim}\;$}}
\nc{\ttbar}{t\bar t}

\newcommand{\fref}[1]{Fig.~\ref{#1}}
\newcommand{\eref}[1]{Eq.~(\ref{#1})}

\newcommand{\cref}[1]{Chapter~\ref{#1}}

\usepackage[utf8]{inputenc}

\begin{document}

\title{Simulating Glueball Production in $N_f = 0$ QCD}

\author{David Curtin}
\email{dcurtin@physics.utoronto.ca}
\affiliation{Department of Physics, University of Toronto, Canada}
 
 \author{Caleb Gemmell}
\email{caleb.gemmell@mail.utoronto.ca}
\affiliation{Department of Physics, University of Toronto, Canada}

 \author{Christopher B. Verhaaren}
\email{verhaaren@physics.byu.edu}
\affiliation{Department of Physics and Astronomy, Brigham Young University, Provo, UT 84602, USA}

\date{\today}

\preprint{PREPRINT}

\begin{abstract}

In an $SU(N_c)$ gauge theory with zero light quark flavours $N_f = 0$, the only hadronic states that form below the confinement scale are composite gluon states called glueballs. These minimal confining sectors arise in many Hidden Valley extensions of the Standard Model, including scenarios that could hold the solution to the dark matter question and the hierarchy problem. Quantitative study of  dark glueball phenomenology  requires an understanding of pure glue hadronization, which to date is severely lacking. In this work we show that significant progress can be made by combining a perturbative pure glue parton shower with a self-consistent and physically motivated parameterization of the unknown non-perturbative physics, thanks to the modest hierarchy between the glueball mass and the confinement scale. We make our simulation code available as the public \texttt{GlueShower} package, the first glueball generator for Hidden Valley theories, and perform preliminary studies of several glueball production observables, with theoretical uncertainties that take the full range of possible hadronization scenarios into account. We hope this will enable new studies of dark sector phenomenology that were previously inaccessible.

\end{abstract}

\pacs{Valid PACS appear here}
\maketitle

\section{Introduction}

The ongoing mysteries of the nature of dark matter (DM) and the electroweak hierarchy problem have long been driving forces for extending the Standard Model (SM). The hierarchy problem motivates searches for new particles below the TeV scale but current experiments have not yet found evidence for their existence. To solve these problems, increasingly interesting and complex dark sectors are being considered. A popular framework that aims to address these issues are Hidden Valley (HV) models \cite{Strassler:2006im}. These often include SM-singlet particles charged under a confining $SU(N_c)$ group, see e.g.~\cite{Kang:2008ea,Bai:2013xga,Renner:2018fhh,Mies:2020mzw}. Couplings to the SM are possible via portal interactions \cite{Holdom:1985ag,Patt:2006fw,Falkowski:2009yz} which tend to be very weak, allowing the possibility of GeV scale states that can evade current experimental bounds. 

Some realizations of the HV framework, generally referred to as neutral naturalness models, solve the little hierarchy problem by cancelling quadratic SM contributions to the Higgs mass with particles uncharged under SM color. Important examples include Mirror Twin Higgs \cite{Chacko:2005pe}, Fraternal Twin Higgs  \cite{Craig:2015pha}, Folded Supersymmetry \cite{Burdman:2006tz}, and many more~\cite{Barbieri:2005ri,Chacko:2005vw,Cai:2008au,Poland:2008ev,Cohen:2018mgv,Cheng:2018gvu}. Additionally, HV models can produce rich and diverse phenomenology, distinct from any SM processes, such as long lived particles (LLPs) \cite{Alimena:2019zri,ATLAS:2013bsk,ATLAS:2019tkk}, soft unclustered energy patterns (SUEPs) \cite{Strassler:2008bv,Knapen:2016hky,Barron:2021btf}, and `dark showers'  \cite{Knapen:2021eip}, leading to semi-visble or emerging jet signatures \cite{Cohen:2015toa,Cohen:2017pzm,Cohen:2020afv,Schwaller:2015gea,Linthorne:2021oiz,CMS:2018bvr,CMS:2021dzg}.


An important special case of Hidden Valleys is the pure Yang-Mills QCD case with $N_f = 0$. The only hadronic states in the dark sector are a spectrum of dark glueballs~\cite{Morningstar:1999rf,Lucini:2008vi,Teper:1998kw,Lucini:2010nv,Athenodorou:2021qvs,Yamanaka:2021xqh}, which can decay to SM states via dimension 6 or 8 operators~\cite{Juknevich:2009gg, Juknevich:2009ji} and have potentially long lifetimes on collider or even cosmological scales. 
$N_f = 0$ QCD-like sectors appear commonly in neutral naturalness models; for example, in the Fraternal Twin Higgs~\cite{Craig:2015pha} only the third generation of SM fermions is mirrored in the dark sector, leaving no strongly interacting states below the confinement scale. Dark glueballs can then be the lightest hadronic states in the twin spectrum. 
Dark glueballs have also been considered as potential DM candidates~\cite{Faraggi:2000pv,Boddy:2014yra,Boddy:2014qxa,GarciaGarcia:2015fol,Soni:2016gzf,Soni:2017nlm,Forestell:2016qhc,Forestell:2017wov,Yamanaka:2019aeq,Yamanaka:2019yek,Jo:2020ggs}, with their relatively strong self-interaction giving rise to interesting astrophysical signatures~\cite{Spergel:1999mh,Weinberg:2013aya}.

Clearly, studying this scenario in detail is highly motivated. However, to date there is no reliably way of simulating dark glueball production in high-energy processes.
Previous studies have resorted to making very simplistic conservative assumptions, like assuming exotic Higgs decays to just two mirror glueballs in studies of LLP signals in neutral naturalness~\cite{Curtin:2015fna,Chacko:2015fbc}.
Another approach is the use of analytical approximations for the final glueball distributions after dark hadronization~\cite{Lichtenstein:2018kno}, but this involves some ad-hoc parameter choices as well as being  inconsistent for all but very high initial energies, due to the relatively high mass of glueballs compared to the confinement scale. 
Clearly, the absence of a reliable event generator for $N_f = 0$ Hidden Valleys severely hampers their phenomenological and experimental study~\cite{Knapen:2021eip}. Our work addresses this shortcoming, opening the door to a large variety of new and detailed investigations.


The difficulty in simulating $N_f = 0$ QCD arises from the unknown nature of hadronization without light quarks. 
For $N_f > 0$ with some dark quark masses below the dark confinement scale, 
 the Lund String model \cite{ANDERSSON198331} can in principle be used to describe hadronization in the dark sector, with the existence of light colored states allowing tubes of color flux to break via light quark pair production.\footnote{Alternative hadronization schemes such as the cluster model \cite{WEBBER1984492} implemented by HERWIG++ \cite{Bahr:2008pv} also only apply for the case with light quarks, as does preconfinement \cite{1979PhLB...83...87A}.}
This is implemented in the Hidden Valley module~\cite{Carloni:2010tw,Carloni:2011kk} of the PYTHIA 8~\cite{Bierlich:2022pfr} generator.
In pure $SU(N_c)$ Yang-Mills theory, on the other hand, no existing hadronization model has been implemented so far. 

In this work we present a simulation strategy for obtaining dark glueball final states from pair produced dark gluons with some initial center-of-mass energy $M$, implemented as the public python package \texttt{GlueShower}\footnote{\href{https://github.com/davidrcurtin/GlueShower}{github.com/davidrcurtin/GlueShower}} for $N_f = 0$ $SU(N_c)$ QCD with $N_c = \{2, \ldots, 12\}$. Because dark gluon production and decay are highly dependent on the specific BSM model, we do not specify those aspects of the dark shower process, instead focusing on the perturbative dark gluon shower and hadronization into dark glueballs.  This can be combined with other event generators for production and decay to give a complete signal simulation for a given dark sector.\footnote{A follow-up paper applying this work to study the possible indirect detection signals from Dark Matter annihilating to dark glueballs in our galaxy is currently in progress \cite{Gemmell}. }

Apart from the practical usefulness of assembling a useable event generator for $N_f = 0$ QCD, the main novelty of \texttt{GlueShower} is our parameterization of different possible hadronization mechanisms, given that the underlying non-perturbative dynamics are even less well understood from first principles than hadronization with light quarks. Our hadronization model is simple and physically motivated, with enough built-in variation to span the space of reasonably possible `jet-like' and `plasma-like' final outcomes.
Despite faithfully incorporating our large theoretical ignorance of pure Yang-Mills hadronization, the resulting predictions are of sufficient precision to make them highly useful for dark sector searches and constraints.

In Sec.~\ref{glueballs}, we briefly review the known properties of glueballs as obtained from lattice QCD. (For simplicity, when referring to gluon and glueballs in this paper we refer to the pure Yang-Mills case, explicitly specifying when we instead refer to SM gluons or states.)
The perturbative aspects of QCD relevant to our Monte Carlo generator are reviewed very briefly for completeness in Sec.~\ref{perturbative shower}. 
Section \ref{hadronization} covers the hadronization process of \texttt{GlueShower}. 
In Sec.~\ref{Simulations} we simulate glueball production for a variety of hadronization assumptions, define a set of 8 benchmark hadronization parameters to span the range of physically reasonable possible outcomes, and make new predictions for observables of glueball production with theoretical uncertainties derived by the variation across these hadronization benchmarks. 
We conclude in Sec.~\ref{conclusion}.

\section{Dark Glueballs}
\label{glueballs}
\begin{figure}[t]
\includegraphics[width=\linewidth]{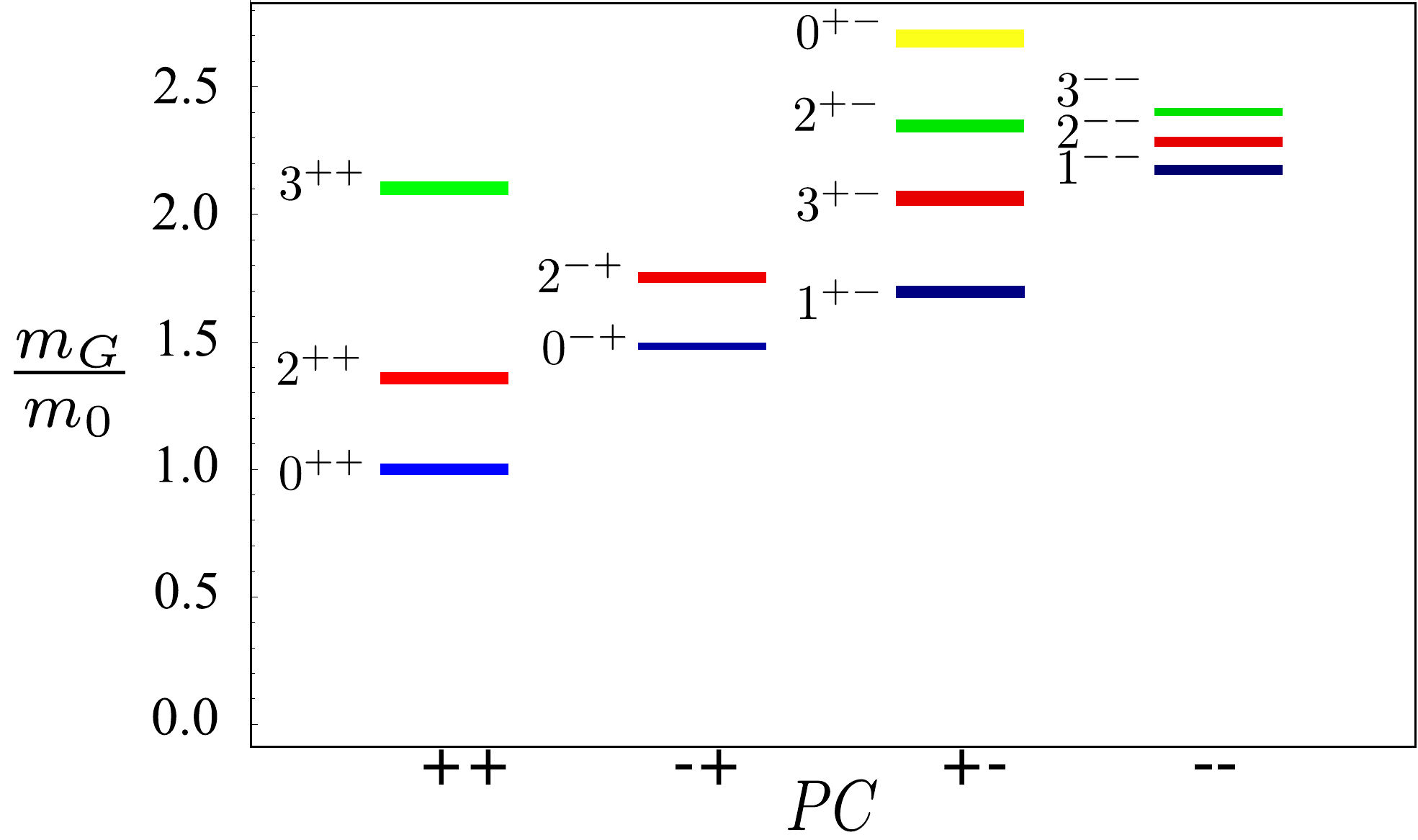}
\caption{Glueball mass $m_G$ spectrum for pure $SU(3)$ Yang-Mills theory \cite{Morningstar:1999rf} in terms of the lightest glueball mass $m_0$, plot taken from \cite{Juknevich:2009gg}.}
\label{fig:glueballspectrum}
\end{figure}

The properties of  $SU(N_c)$ glueballs have been studied on the lattice for decades~\cite{Morningstar:1999rf,Lucini:2008vi,Teper:1998kw,Lucini:2010nv,Athenodorou:2021qvs,Yamanaka:2021xqh}, establishing a spectrum of twelve stable states in the absence of external couplings, as shown in Figure \ref{fig:glueballspectrum}. These states are distinguished by their $J^{PC}$ quantum numbers, and their masses can be parameterised entirely in terms of the confinement scale, $\Lambda$.

In this work we use the lattice values calculated in~\cite{Athenodorou:2021qvs}. Across the range of $N_c=\{2,\ldots,12\}$ values we consider, the lightest glueball mass $m_0$ is approximately 6$\Lambda$. This is much heavier than e.g. SM hadrons, and important for our characterization of glueball hadronization. Additionally they also provide the glueball masses for the next nine states and relate the string tension $\sqrt{\sigma}$ to the three-loop confinement scale in the $\overline{MS}$ scheme.
Since $\Lambda_{\overline{MS}}$ is quoted to 3-loop accuracy we also use the 3-loop running coupling as given in \cite{Prosperi:2006hx}, which has been used with tree-level splitting functions in other Monte Carlo generators \cite{Bahr:2008pv}.\footnote{Using the 3-loop running coupling is simply to be consistent with the lattice results and not indicative of a higher level of accuracy, as any improvement compared to the one-loop result is subdominant to the uncertainties of our hadronization model.}
These values are provided for $N_c$ between 2 and 12, and because the perturbative QCD calculations are easily adjusted for a general $N_c$, \texttt{GlueShower} generates glueball showers for this same range of $SU(N_c)$ confining sectors. This opens the door for simulating the phenomenology of a wide range of interesting exotic dark sectors, and hopefully encourages study of confining dark sectors beyond the SM-like $SU(3)$ case \cite{Boddy:2014yra,Soni:2016gzf,Batell:2020qad,Kilic:2021zqu}.

Glueball wavefunctions have also been studied on the lattice and by other methods. These have determined the average size of glueballs, with the $r_\text{rms}$ value for the $0^{++}$ glueball typically found to be of order $\sim\Lambda^{-1}$ \cite{Hou:2001ig,Ishii:2001am,Loan:2006gm}. The next heaviest state, $2^{++}$, is approximately twice the size of the $0^{++}$ state \cite{Loan:2006gm}. There is still uncertainty in these measurements and questions regarding whether quenched QCD lattice studies will agree with the pure Yang-Mills results. Still, these small radii support focussing on local physics rather than considering glueballs as truly extended objects, which guides our discussion of hadronization below.

\section{Perturbative Shower Review}
\label{perturbative shower}
Perturbative QCD is an extensively studied and established field within quantum field theory. Following~ \cite{Ellis:1991qj,Sjostrand:2006za} we briefly review the salient details of perturbative parton showers in this section for completeness and to establish notation. More complete documentation of how this is implemented in \texttt{GlueShower} is found in Appendix \ref{app:pQCD}.

For $N_f = 0$ HV models, the only parton is the gluon. Our code works in the centre-of-mass frame of a two-gluon initial state with  invariant mass $M$.\footnote{This is the only initial state we consider, as it is of the most use to BSM physics studies. We leave the study of other interesting cases, such as a gas of gluons that is cooled until it undergoes confinement, for future investigations.} This amounts to simulating gluon production from the decay of a massive scalar particle of mass $M$, but can be easily generalized to, for example, direct di-gluon production via effective operators or intermediate states. These initial gluons are produced with large virtualities (effective mass-squared) $t$, and as they split into more gluons produce an increasing ensemble of lower virtuality gluons. The energies and virtualities of these evolving gluons are described by perturbative QCD and can be simulated as a parton shower.

 The probability that a gluon splits into two gluons with energies $z$ and $(1-z)$, where $z$ is the energy fraction of the mother dark gluon, is determined only by the gluon-to-gluon splitting function:
\begin{equation}
P_{gg}(z) = 2C_A\bigg[\frac{z}{1-z}+\frac{1-z}{z}+z(1-z)\bigg]~,
\end{equation}
where $C_A = N_c$. This splitting function is also used to define the Sudakov form factor, which gives the probability that the gluon evolves from an initial virtuality, $t_0$, to a lower virtuality, $t$, without splitting:

\begin{equation}
\Delta(t) = \text{exp}\bigg[ -\int_{t_0}^{t} \frac{dt'}{t'} \int dz \frac{\alpha_s}{2\pi} P_{gg}(z) \bigg]~.
\end{equation}

Note that this Sudakov form factor only accounts for the  leading-log collinear gluon enhancements, which is unable to reproduce the correct scaling of parton multiplicity with centre of mass energy. However, this is easily rectified in standard implementations of parton showers by imposing angular ordering on subsequent splittings, which accounts for soft gluon interference effects at leading order in $1/N_c^2$ \cite{Marchesini:1983bm}, see Appendix \ref{app:pQCD} for further details.

Monte Carlo parton shower evolution can be framed in the following way: Given a gluon with virtuality and energy, $(t_1,z_1)$, after some step in the evolution, what is its new virtuality and energy, $(t_2,z_2)$? Note that a gluon can only decrease its virtuality by splitting. Thus, finding a value for $t_2$ implies the gluon split in the intermediate step. To calculate $t_2$ a random number $\mathscr{R}\in[0,1]$ is generated and $t_2$ is found by solving

\begin{equation}
\Delta(t_2) = \frac{\Delta(t_1)}{\mathscr{R}}~.
\end{equation}

If there is no solution, then the gluon does not split. In this case the gluon can only hadronize, and its shower terminates. If a $t_2$ is found, $z_2$ is determined by generating another random number $\mathscr{R}'\in[0,1]$ and solving:

\begin{equation}
\int_{z_{\text{min}}}^{z} dz' \frac{\alpha_s}{2\pi} P_{gg}(z') = \mathscr{R}' \int_{z_{\text{min}}}^{1 - z_{\text{min}}} dz' \frac{\alpha_s}{2\pi} P_{gg}(z'),
\end{equation}
where $z_\text{min}$ is set by kinematic thresholds of the possible gluon splittings. This Monte Carlo generation is implemented in \texttt{GlueShower} as detailed in Appendix \ref{app:pQCD}. We also note that since $\alpha_s \propto N_c^{-1}$ at 1-loop order, and the perturbative shower is only dependent on $\alpha_s P_{gg}(z)$, changes to $N_c$ only affect the shower due to the slight dependence of glueball masses on $N_c$. In our numerical studies below we therefore only show the $N_c = 3$ case, with other numbers of colours giving similar behaviour.

\begin{figure*}[t]
\includegraphics[width=0.7 \textwidth]{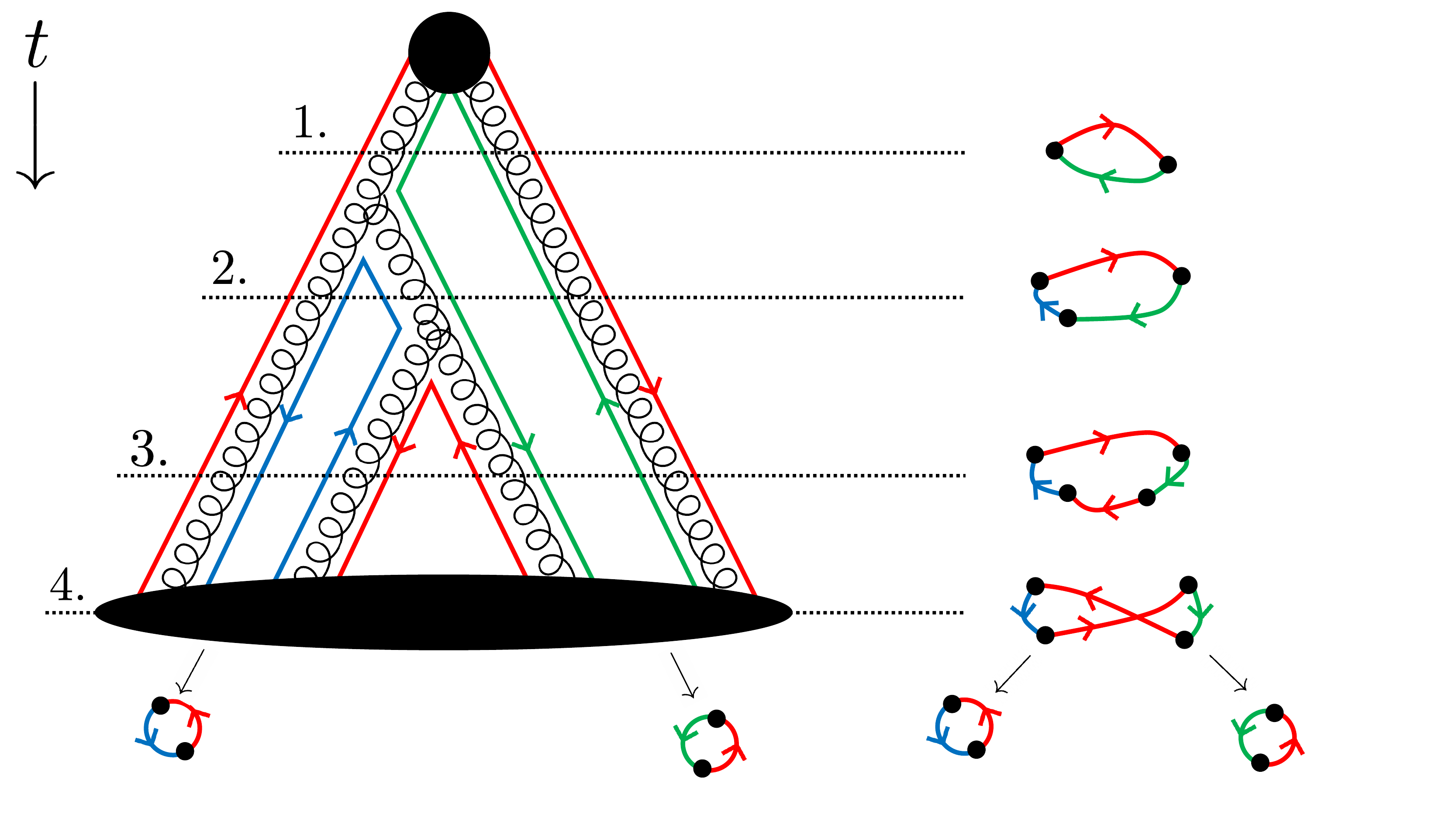}
\caption{Comparison between Feynman diagram-like (left) and flux string-like (right) cartoons of the evolution of colour singlet separation in $N_f = 0$ QCD.}
\label{fig:glue diagrams}
\end{figure*}

Lastly, as is standard in parton shower MC algorithms, we work in the leading colour limit, which  is equivalent to taking $N_c \rightarrow \infty$ with $\alpha_sN_c$ kept constant \cite{tHooft:1973alw}. This amounts to ignoring higher order colour interference effects in $1/N_c^2$ expansions. 
Additionally in this limit we can use the simple representation of gluons being the direct product of a fundamental and anti-fundamental, since the weight of the singlet in $N_c \otimes \overline{N_c} = (N_c^2 - 1) \oplus 1$ vanishes. It is in this limit that the t' Hooft double line notation can be used to trace colour flow, as shown in Fig. \ref{fig:glue diagrams}. From SM QCD we know this approximation works well for $N_c = 3$, and thus also for $N_c > 3$. 
It is possible that this approximation introduces larger errors for $N_c = 2$, but we leave the problem of including the subleading colour corrections for future investigations.

\section{Hadronization}
\label{hadronization}

While the perturbative aspects of the parton shower have been extensively studied, the non-perturbative aspects of QCD have not been computed from first principles. 
We follow the usual approach, outlined in the previous section, whereby we iteratively evolve gluons from their initial virtuality and undergo splittings above some hadronization scale $\Lambda_\text{had}$, which is typically taken to be of the same order as the confinement scale $\Lambda$ (though it turns out this assumption must be modified for our case).
Some hadronization model capturing the non-perturbative dynamics of glueball formation is then needed to turn the gluon final states of the shower into physical final states. 
In SM QCD, phenomenological hadronization models \cite{ANDERSSON198331,WEBBER1984492} can be tuned to reproduce the observed data, but these rely on the existence of light quarks, which either allow a flux tube string to fragment, or allow colour singlet clusters to form during preconfinement~\cite{1979PhLB...83...87A}. 
For pure Yang-Mills, the corresponding hadronization models have not been formulated and at any rate cannot be tuned to data prior to the discovery of a corresponding dark sector.\footnote{A highly exciting possibility is understanding pure glue hadronization from first principles on the lattice. This is very challenging, but there has been significant recent progress for $N_f >0$, see e.g.~\cite{Bulava:2019iut}.} We therefore provide a novel parameterization of the various possibilities for $N_f = 0$ QCD hadronization, relying on simple physically motivated arguments to capture the range of both `jet-like' or `plasma-like' possibilities for how pure glue might hadronize, producing a consistent spectrum of simulated glueball final states obeying full energy-momentum conservation. This accurately captures present theoretical uncertainties while allowing for the quantitative study of pure glue hidden valleys with sufficient precision for searches and constraints. 

\subsection{An Intuitive Cartoon}

We begin with a representative cartoon to guide our intuition for pure glue hadronization. In this cartoon we assume there is only one glueball species and consider the simple example of producing two gluons that eventually hadronize into two glueballs. Because the glueballs are colour-singlet final states, and we begin with two back-to-back colour-octet gluons, long-range colour exchange must occur at some point along the shower/hadronization \emph{if the center-of-mass energy is high enough}.\footnote{The minimum initial energy required for separation into multiple colour singlets will be made more precise in the next subsection.}
Thus, before the glueballs can form, at minimum one of the original gluons must split into three gluons, with one new gluon joining the other branch and allowing colour singlet states to form, referred to as colour octet neutralisation \cite{Minkowski:2000qp}.\footnote{Other detailed splitting histories are of course possible but do not change our qualitative argument.} This illustrative cartoon is depicted in a Feynman-diagram-like way on the left side of Fig.~\ref{fig:glue diagrams}, showing gluon splittings and flow of colour charge.

Alternatively, one can consider how tubes of colour flux in the fundamental representation evolve throughout the shower. This equivalent representation is depicted on the right side of Fig.~\ref{fig:glue diagrams}. The initial state is an overall colour singlet loop, with the gluons behaving as localised energy or `kinks' in the loop. 
Forming two final state glueballs can only be accomplished by the initial loop fragmenting in two. As the loop evolves, gluon splitting introduces new kinks and new colors for the flux tube segments in the loop.
Towards the end of the shower, step 3, enough gluons have formed to allow two flux tube segments in the loop to have the same colour, making fragmentation into two colour singlet glueballs possible by crossing the same-colour flux-tube segments. \emph{Thus, we see that glueball hadronization can be pictured as crossing color-fundamental flux tubes `pinching off' one flux tube loop into two separate loops, which becomes possible even after a small number of gluon splittings.}

The tube crossing picture is supported by the fact that loop fragmentation is hugely energetically preferred once it is possible, since flux tube crossing has little energy cost compared to the energy stored in flux tubes of length $\gg \Lambda^{-1}$. Making the crossing possible only requires the exchange of arbitrarily soft IR gluons carrying colour information, which carries no particularly relevant suppression and is in fact enhanced by $\alpha_s$ running to large values at low energies. This is also consistent with glueball radii being of order the confinement scale, suggesting their formation is dominated by local physics. 

We can compare the glueball flux tubes with the Lund String Model used for QCD. For the simple case of quark pair production, the shower and subsequent hadronization process essentially follows the evolution of an open flux tube with the quarks at either end. These flux tubes are easily broken through light quark pair production, which increases the open flux tube multiplicity. The final state flux tube states with quarks and antiquarks on either end are then associated with a variety of mesons. $N_f > 0$ QCD is predominately determined by the fragmentation of open strings, while the final state for $N_f = 0$ QCD HV theories is  determined by closed loops pinching and fragmenting.

In summary, formation of individual glueballs requires crossing flux tubes, which seems to be entirely unsuppressed once a small number of (possibly very soft) gluons have been exchanged between different branches of the shower. This required gluon exchange in turn becomes  unsuppressed once the virtualities in the shower approach the confinement scale.  
\emph{This motivates the intuition that glueball formation should occur around the confinement scale, qualitatively similar to SM jet-like behaviour.}

\subsection{\texttt{GlueShower} Hadronization Implementation}

We now describe our pure glue hadronization algorithm in detail, first for the default jet-like assumption as motivated above, and then for more exotic plasma-like behavior, to cover all physically reasonable possibilities.

\subsubsection{Jet-like glueball hadronization} 

The assumption that glueball production proceeds in a qualitatively similarly jet-like fashion as for SM QCD is well-motivated by physical arguments, as explained in the previous section. However, the detailed question still remains how to map some arrangement of gluons at the end of the perturbative shower to a set of final state glueballs. One could implement a full Lund-like string model of closed flux tubes and their crossing, and vary its parameters to obtain a range of possibilities for glueball hadronization. This would be quite involved, and we leave this for future investigations. However, thanks to the significant separation between the confinement scale $\Lambda$ and the lightest glueball  mass $m_0$, we can already make great progress with much simpler physical arguments. 

We begin by assuming that there is just a single glueball species of mass $m_0$, and consider a gluon in the perturbative shower that has virtuality $\sqrt{t} > 2 m_0$ and does not undergo further  splittings above scale $2 m_0$. If the gluon were allowed to evolve further down in virtuality, any subsequent splittings cannot result in more than one on-shell glueball unless there is significant momentum exchange with other branches of the shower. Instead, subsequent splittings result in lower-energy gluons physically clustered around the original gluon 4-momentum in a decay cone that is at most as wide as one arising from the decay of a particle with mass below $2 m_0$. As $\sqrt{t} \to \Lambda$, hadronization must combine these soft gluons (plus some soft IR gluons to exchange colour information with other branches of the shower) into an on-shell, colour-singlet glueball with mass $m_0$. 

Therefore, we argue that for a given shower history, an \emph{upper bound} on the number of glueball states produced can be obtained by setting the hadronization scale to $\Lambda_\text{had} = 2 m_0$ and simply turning the gluons at the end of the shower into on-shell glueballs of mass $m_0$.
Note that $2 m_0 \sim 12 \Lambda$ is well within the perturbative regime, so the simulated momenta of  gluons at the end of the shower are highly reliable. 
However, in converting these gluons to glueballs some soft gluons must be exchanged with other branches of the shower to form colour singlets. This suggests momentum transfers of order $\Lambda/2 m_0$, but the relatively high mass of glueballs makes this correction factor smaller than $\approx 10\%$. Therefore, naively turning gluons with $\sqrt{t} = \Lambda_\text{had}$ into on-shell glueballs is likely to be a good approximation.\footnote{This method brushes over the specifics of how colour information is exchanged to create colour singlets, reminiscent of early independent fragmentation models \cite{FIELD19781}. However, these early models proved to be historically useful in SM QCD, are even more useful for pioneering BSM studies in our case due the larger mass of the glueball hadrons in $N_f = 0$ compared to the typical momentum transfer involved in colour exchange.}

How could this upper bound on glueball multiplicity be violated? First,  it may be possible for  two neighbouring branches of the shower to exchange gluons with momenta between $2 m_0$ and $\Lambda$, allowing two gluons with virtuality $2 m_0$ each to result in three instead of two final state glueballs. This can still be regarded as a (marginally) perturbative process in the shower's regime of validity. On the other hand, this assumption can be violated entirely in the non-perturbative regime if the flux strings between two branches contain enough energy to produce a third glueball.

We first discuss the perturbative possibility, where the size of the momentum exchange gives us some hope of using the perturbative shower and simple phase space arguments to estimate this rate $P_{2\to3}$ of turning two gluons at the end of the shower into three glueballs. We  overestimate $P_{2\to3}$ to show that it is small enough to ignore at our current level of precision. 

Consider two gluons 1 and 2 at the end of the shower, both with virtuality $2 m_0$, energies $E_{1,2}$ and angle $\theta$ between their momenta. To overestimate $P_{2\to3} = P_{2\to3}(E_1, E_2, \theta)$, we assume that a third glueball is formed if the gluons split $1 \to 1'3$ and $2 \to 2' 4$ such that $m_{1', 2'} > m_0$, the daughters have sufficient invariant mass to form the glueball $m_{34} \geq m_0$ and the two daughter momenta are close in phase space compared to the confinement scale, i.e. $\Delta p_{34} \equiv |\vec p_3 - \vec p_4| < a \Lambda$ for $a \sim \mathcal{O}(1)$, to enable the merger. We choose $a = 2$ but the precise value does not significantly affect our result. 
We can therefore estimate
\begin{eqnarray}
P_{2\to3}(E_1, E_2, \theta) &=& \int d t_1 d t_2 d z_1 d z_2
\\ \nonumber && 
P_{1\to 1'3, 2\to 2'4, m_{1',2'} > m_0} (E_1, E_2, \theta)   \\
\nonumber &&
P_{m_{34} > m_0, \Delta p_{34} < a \Lambda} (|p_3|, |p_4|, \theta)
\end{eqnarray}
This integrates over all possible splittings $(t_1, z_1), (t_2, z_2)$ of the two parent gluons that produce daughters 3 and 4. The first term (splitting probability) just evaluates the Sudakov to give the splitting probability of both parent glueballs producing two daughters with momentum $p_3, p_4$ such that the parents still have sufficient virtuality to form their own glueballs. The second (merger probability) term is the probability, given a random emission angle for each daughter in the transverse plane of the parent momentum, that the two daughters could in principle combine to form a third glueball according to our above criteria.

\begin{figure}[t]
\includegraphics[width=\linewidth]{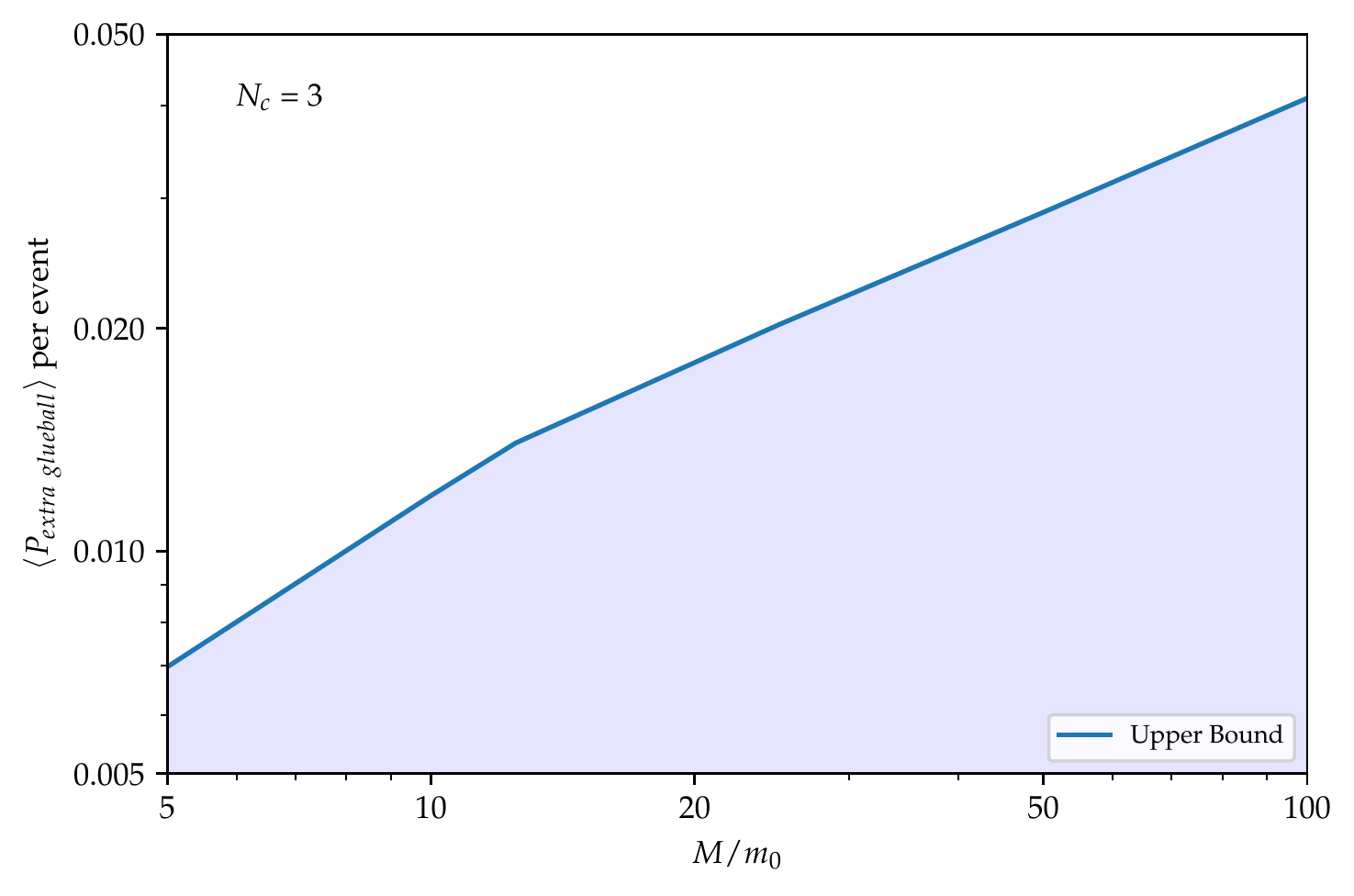}
\caption{Upper bound on the probability of producing more glueballs of mass $m_0$ in gluon pair production events than the naive upper multiplicity bound obtained by terminating the pure glue shower at $\Lambda_\text{had} = 2 m_0$, as a function of the center-of-mass energy $M$.}
\label{fig:extra glueball}
\end{figure}

To simplify evaluation of this integral, we overestimate both terms separately. The splitting probability (first term) is overestimated by letting both gluons 1 and 2 run from $\sqrt{t} = 2 m_0$ down to $\sqrt{t} = 2 \Lambda$, allowing for emission of daughters with virtualities as low as $\Lambda$. This is a huge overestimate since we are not enforcing $m_{1', 2'} > m_0$, and turns $P_{1\to 1'3, 2\to 2'4, m_{1',2'} > m_0} (E_1, E_2, \theta; p_3, p_4)$ into a constant $P_\text{split}^\text{max}$ that sits outside the integral.
We then set the virtualities of the daughters $3,4$ to be the largest possible masses allowed by a given splitting to define their 4-momenta $p_3, p_4$, which in turn maximizes the merger probability (second term). This allows us to define
\begin{eqnarray}
P_{2\to3}^\text{max} (E_1, E_2, \theta) &\equiv& P_\text{split}(E_1, E_2) \times 
\\ \nonumber &&  \int d t_1 d t_2 d z_1 d z_2 P_\text{merge}(|p_3|, |p_4|, \theta)
\end{eqnarray}
satisfying $P_{2\to3}^\text{max}  > P_{2\to3}$. Having thus obtained a function $P_{2\to3}^\text{max} (E_1, E_2, \theta)$, it is then straightforward to generate events using the perturbative pure-gluon shower for a variety of initial center-of-mass energies $M$, terminate at hadronization scale $\Lambda_\text{had} = 2 m_0$, and evaluate the chance of obtaining an extra glueball compared to our naive expectation of turning gluons with virtuality $2 m_0$ into glueballs:
\begin{equation}
P_\mathrm{extra\ glueball} = 1 - \prod_{i,j} (1-P_{2\to3}^\text{max} (E_i, E_j, \theta_{ij}))
\end{equation}
where the product is over all gluon pairs $(i,j)$ in the event. Averaged over all events, this probability is shown in \fref{fig:extra glueball} for $N_c = 3$, but the result is nearly identical for other  numbers of colours. As expected, the chance of producing additional glueballs beyond our upper bound increases with center-of-mass energy since this gives more gluons and more chances for the required splittings and mergers. However, for $M \lesssim 100 m_0$, the error introduced by ignoring $2\to3$ production of glueballs is smaller than 5\%, and we are justified in ignoring it for our current implementation. 

We now consider non-perturbative effects that could produce additional glueballs, orienting ourselves in the SM analogues of Schwinger pair production in QED~\cite{PhysRev.82.664}, or pion production in the ``snapping'' of colour-fundamental strings~\cite{PhysRevD.20.179}. 
The latter case is of most physical interest. The energy density of the colour string is given by the string tension $\sigma$, and the production rate of hadrons in string fragmentation scales as
\begin{equation}
P_\mathrm{hadron} \propto \text{exp}\bigg[-\frac{\pi m_\mathrm{hadron}^2}{\sigma}\bigg] , 
\end{equation}
since the constituent masses of the partons have to be provided by the colour background field. 
Ignoring any additional suppressions that may arise from the required geometrical arrangement of two colour flux tubes to allow for the creation of a glueball, the fact that the glueball mass is much larger than the mass of SM pions compared to the string tension should result in a large suppression on the number of glueballs produced from the dynamics of the non-perturbative colour strings. For $N_f = 0, N_c = 3$, the ratio of the exponential factors is
\begin{equation}
\frac{P_\mathrm{GB}}{P_\mathrm{\pi}} \sim 10^{-16} \ .
\end{equation}
where the numerator was evaluated using $\Lambda_{\overline{MS}} / \sqrt{\sigma} = 0.5424$ and the lightest glueball mass, $m_{GB} = 6.28 \Lambda_{\overline{MS}}$~\cite{Athenodorou:2021qvs}. Note that for different $SU(N)$ groups, the constants change slightly but the extreme suppression persists. We therefore conclude that both perturbative and non-perturbative effects should not invalidate our jet-like approach.


If setting $\Lambda_\text{had} = 2 m_0$ and simply turning final state gluons into on-shell glueballs gives the largest possible glueball multiplicity per event, how can we take into account the possibility that the real number of produced glueballs could be lower? 
Physically, this would correspond to colour singlets forming via gluon exchange at a higher virtuality scale than naively expected, which is certainly a possibility given the unknown details of non-perturbative $N_f = 0$ QCD. We can obtain consistent events representing this scenario by simply terminating the shower at a higher scale before turning final state gluons into on-shell glueballs. In other words, we set
\begin{equation}
\Lambda_\text{had} = c \times 2 m_0
\end{equation}
where $c > 1$ is a dimensionless parameter encoding our assumption of the higher scale $\Lambda_\text{had}$  where the shower fragments into disconnected colour singlets that each yield one glueball. Varying $c \sim \mathcal{O}(1)$ gives us a controlled way to parameterize different possible assumptions on jet-like glueball hadronization, and explore the effect of this uncertainty on observables while still producing fully consistent shower histories that conserve energy and momentum for arbitrary choices of $c \geq1 $ and $M/m_0 > 2$.

So far, we have only considered a single glueball species, but in reality there is a spectrum of roughly a dozen different glueball states with different $J^{PC}$ quantum numbers. 
While our arguments support the notion that the \emph{inclusive} glueball \emph{multiplicity} and \emph{momentum distributions} are dominated by local jet-like physics, the same is not true for \emph{exclusive} distributions for each glueball species. 
Soft gluon exchange at the scale $\sim \Lambda$ (also the scale of mass differences between glueball species) can easily exchange angular momentum and other quantum numbers to turn a given candidate gluon into a variety of different glueball species, and rigorously analyzing these non-local effects is far beyond our scope. 
We therefore adopt a much simpler approach of assuming that the relative multiplicities of different glueballs follows a thermal distribution~\cite{Falkowski:2009yz} in the absence of other threshold effects, with the probability for producing glueball state $J$ given by
\begin{equation}
\label{e.PJ}
P_J \propto  (2J + 1)\bigg(\frac{m_J}{m_0}\bigg)^{3/2}e^{-(m_J - m_0)/T_\text{had}}~,
\end{equation}
where the glueball masses for different $N_c$ are known from the lattice, and we define a hadronization temperature
\begin{equation}
T_\text{had} = d \ T_c
\end{equation}
which is related by a dimensionless coefficient $d \sim \mathcal{O}(1)$  the critical temperature of the $N_f = 0$ QCD phase transition $T_c$.
This is justified by investigations of final state distribution from closed string emission  \cite{Manes:2001cs} that support a thermal model, where $T_c$ is taken to be the Hagedorn temperature. Consequently, we assume the Hagedorn temperature is the critical temperature of deconfinement \cite{Blanchard:2004du,Noronha-Hostler:2010nut}. 
In SM QCD, the critical temperature is 150 MeV \cite{Petreczky:2012rq}, smaller than the confinement scale. In $N_f = 0$ Yang-Mills theories the relation between the critical temperature and string tension has been studied on the lattice \cite{Lucini:2003zr, Boyd:1996bx, Lucini:2005vg, Lucini:2012wq}, with \cite{Lucini:2012wq} finding the relation
         \begin{equation}
    \frac{T_c}{\sqrt{\sigma}} = 0.5949 + \frac{0.458}{N_c^2}~,
    \end{equation}
giving a critical temperature that is slightly larger than the confinement scale. By combining this result with the three-loop relation between the confinement scale $\Lambda$ and the string tension, the relative glueball multiplicities are entirely determined by the number of colours $N_c$ and the nuisance parameter $d $.

We incorporate these probabilities into our hadronization routine in the following manner. Once it is determined that a gluon can no longer split during the perturbative shower, it remains with some virtuality  $\sqrt{t} \geq \Lambda_\text{had} = 2 c m_0$. This gluon then selects a random glueball final state with on-shell mass below its current virtuality, weighted by the probabilities in \eref{e.PJ}.
This treats the thermal probabilities $P_J$ as fundamental, and introduces  some additional threshold effects that favour light glueball production, since a high-virtuality-gluon that does not split before termination of the shower can have more glueball final states kinematically accessible to it than a gluon that was produced during the shower with virtuality close to $2\, c\, m_0$. The relative multiplicity of glueball species produced by this hadronzation routine will therefore skew towards lighter flavours than the thermal $\{P_J\}$ alone, but we believe this is a physically reasonable prediction of the perturbative shower based on little more than phase space arguments. 

Additionally, colour rope hadronization \cite{Biro:1984cf} in SM QCD can affect the final state relative multiplicities by enhancing strangeness production \cite{Bierlich:2017sxk} via an increased string tension. This could be relevant for the $N_f = 0$ QCD case since the two fundamental flux strings between gluons would behave collectively when this effect is included, changing the relative multiplicity of glueballs. Including these dynamics is beyond our scope, but we account for their possible effect by not treating $T_\text{had}$ as firmly determined, and vary $d$ to parameterize the theory uncertainty in relative glueball species multiplicity. 
Higher hadronization temperatures favour production of heavier glueball states. Depending on the full details of the theory and the operator by which glueballs decay to SM states~\cite{Juknevich:2009gg, Juknevich:2009ji}, this can significantly affect the visible phenomenology, and it is important to treat $d \sim \mathcal{O}(1)$ as a nuisance parameter in quantitative analyses. We demonstrate this by studying some benchmarks below. 

While we expect our approach of randomly assigning glueball species identity based on thermal probabilities to be reasonable on average over many simulated events, this simple approximation is unlikely to give accurate intra-event correlations between separations in momentum space and relative species assignments of different glueballs, for example. Even so, this simple parameterization should be sufficient for many first phenomenological investigations. 


\subsubsection{Plasma-like glueball hadronization} 

While the jet-like showering and hadronization behaviour is highly physically motivated, we want \texttt{GlueShower} to cover the largest range of physically possible showering behavior. We therefore also consider a much more exotic plasma-like or SUEP-like regime. 

Within the jet-like assumption, the fragmentation of closed flux tubes is viewed as an IR process that results in the immediate formation of on-shell glueballs. However, if fragmentation into colour singlets occurs at a significantly higher scale, then a population of high mass closed flux tubes could be produced.\footnote{We thank Matthew Strassler for bringing this possibility to our attention.} These states could be treated as a collection very excited glueballs, or a hot ball of gluon plasma. Much like a quark-gluon plasma that evaporates via quasi-isotropic emission of pions, this pure glue plasma would evaporate by emitting glueballs approximately isotropically with thermal momenta in its restframe. 

To implement this possibility within \texttt{GlueShower}, we introduce boolean parameter \texttt{plasma\_mode} which is \texttt{False} by default (jet behaviour) but can be set to \texttt{True} to enable plasma behaviour. In plasma mode, the shower still terminates at a scale determined by setting $c > 1$ just as in jet mode,
\begin{equation}
\Lambda_\text{had} = c \times 2 m_0 , 
\end{equation}
but this scale is now interpreted as the scale below which the shower separates into singlet plasma balls, and  each final-state gluon is turned into a singlet of mass $m_\text{plasma} = c\, m_0$ instead of a glueball $J$ of mass $m_J$. 

The evaporation of these plasma balls into glueballs is treated analogously to dark hadron production in SUEP scenarios~\cite{Strassler:2008bv,Knapen:2016hky,Swisdak:2013vxa}, assuming isotropic thermal glueball emission. 
We assume that the thermal glueball energy distribution and their relative species probabilities are dictated by the same temperature $T_\text{had} = d \cdot T_c$, see \eref{e.PJ}. 
We borrow the SUEP-simulation methods used in \cite{Knapen:2016hky,Swisdak:2013vxa} to generate isotropic glueball momenta in the restframe of each plasma. In short, isotropic glueball momenta are successively generated with a thermal energy distribution until additional gluon emission would cause the total invariant mass of all gluon momenta to exceed the plasma mass $m_\text{plasma}$. For each glueball, its species is randomly picked weighted by the thermal probabilities $P_J$ in \eref{e.PJ}. 
Once glueball emission is completed for a given plasma-ball, the entire system of daughter-glueballs is slightly boosted and its kinetic energies rescaled to exactly equalize their rest frame and invariant mass with the original plasma ball. 

In plasma mode, glueballs are therefore produced in a manner that lies between the jet-like behavior of QCD and the purely isotropic behavior of pseudo-conformal theories that realize the SUEP scenario: the shower generates branches of total invariant mass $m_\text{plasma} = \Lambda_\text{had}/2$, which become plasma balls that evaporate via isotropic thermal glueball emission in their respective restframes. 

We emphasize that this possibility is highly exotic: we regard jet-like hadronization as far more physically motivated, since realizing plasma-like behavior requires very unusual long-distance non-perturbative effects that arrest further  fragmentation well within the naive regime of reliability for the perturbative shower. However, we include it in our code to make sure that even highly exotic hypotheses for the non-perturbative dynamics of $N_f = 0$ QCD can be qualitatively accommodated.

\subsubsection{Input Parameters for \texttt{GlueShower}}

In summary, \texttt{GlueShower} has two purely physical parameters: $N_c \in (2,3,\ldots 12)$ to specify the dark gauge group and $m_0$ to specify the mass of the lightest glueball.\footnote{The large $N_c$ regime can be well-covered by simulating $N_c = 12$.} This determines the confinement scale $\Lambda$, critical temperature  $T_c$, and the 3-loop running coupling $\alpha_S(\mu$). 
For a given run, one also specifies the initial center-of-mass energy $M$ for the di-gluon initial state (corresponding to di-gluon pair production in the decay of a scalar particle of mass $M$) as well as the number of shower histories to simulate.
Theoretical uncertainties of dark hadronization are captured in two-and-a-half nuisance parameters: the boolean parameter \texttt{plasma\_mode} which specifies whether gluons hardronize in the physically motivated jet-like or the more exotic plasma-like fashion, and multiplicative parameters $c = \Lambda_\text{had}/(2 m_0)$ and $d = T_\text{had}/T_c$ which set the hadronization/singlet formation scale and the hadronization (and plasma-ball, if in plasma mode) temperature respectively.

\begin{figure}[t]
\includegraphics[width=\linewidth]{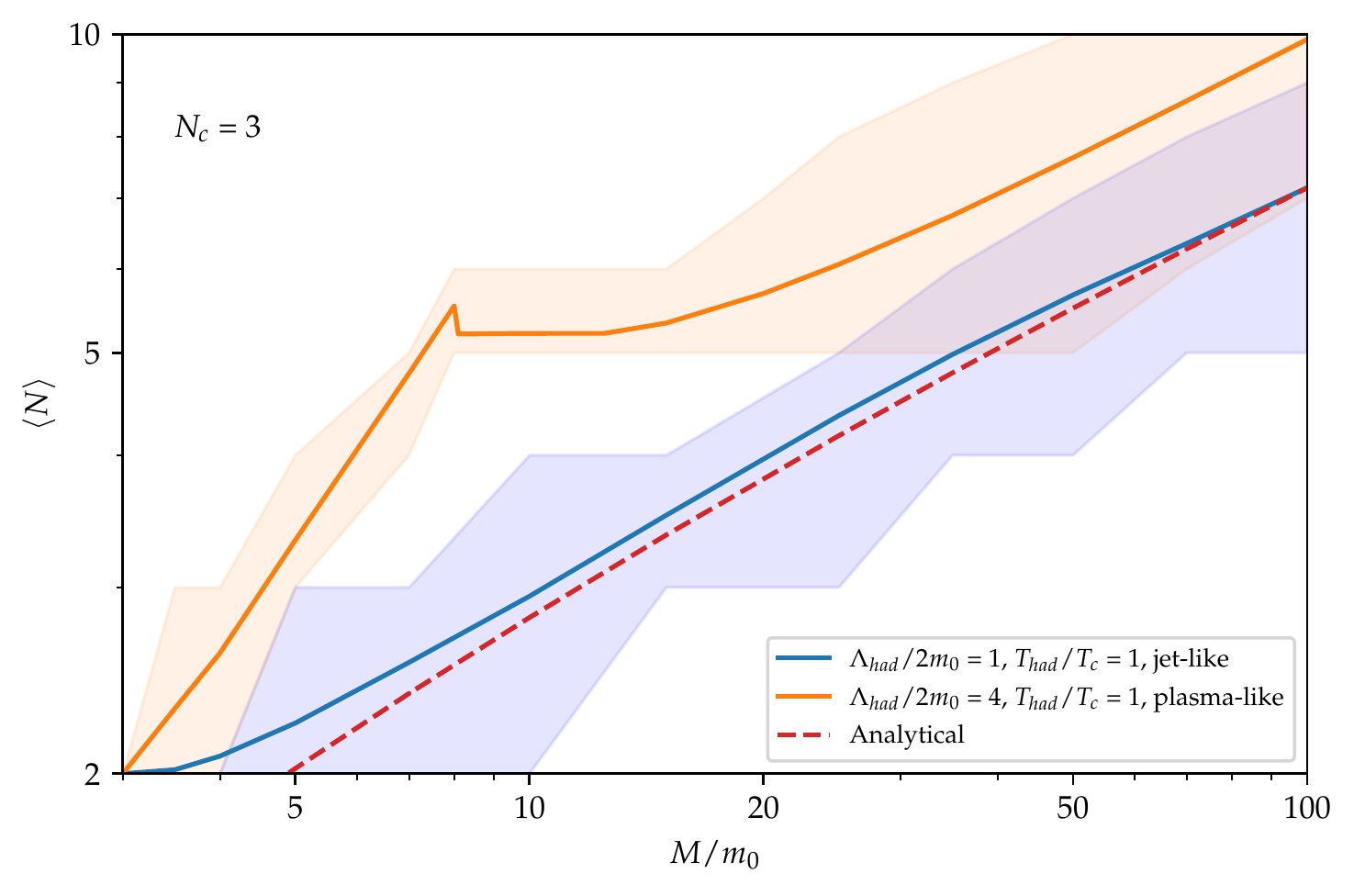}
\caption{
Average glueball multiplicity  $\langle N \rangle$ for $N_c = 3$ as a function of center-of-mass energy $M$ relative to the lightest glueball mass $m_0$. 
The jet-like case with the highest possible glueball multiplicity is shown in blue, a representative plasma-like case in orange (see legend). 
The bands shows the $16\%$ to $84\%$ percentile range of multiplicities in the event distributions for each $M/m_0$ (jagged since multiplicities are small integers).
The analytical expectation of \eref{mult eqn}, normalized to the $M/m_0 = 100$ jet-like simulation, is shown as the dashed red line.
}
\label{fig:analytical multiplicity}
\end{figure}

\section{Simulation of dark glueball final states}
\label{Simulations}
In this section we study events generated by \texttt{GlueShower}. We first explore the basic jet-like shower case, using parameters \texttt{plasma\_mode = False}, $ \Lambda_\text{had}/(2 m_0) = c=1$, and $T_\text{had}/T_c = d=1$ to demonstrate the code and its output, as well as comparing  the multiplicity scaling and shape of fragmentation functions to analytical approximations in their expected regime of validity.
We then comment on the qualitative differences in the plasma-like case, using parameters \texttt{plasma\_mode = True}, $ \Lambda_\text{had}/(2 m_0) =4$, and $T_\text{had}/T_c =1$.
Finally, we argue that a set of 8 benchmark values for the nuisance parameters covers the physically motivated range of hadronization possibilities for glueball production (\texttt{plasma\_mode = FALSE \{TRUE\}}, $c = 1,2$ $\{4,6\}$, $d = 1,2$) and study the resulting range of physical predictions for some observables.

\subsection{Jet-like Hadronization}
We first demonstrate how the inclusive glueball multiplicity scales with initial centre-of-mass energy $M$ in Fig.~\ref{fig:analytical multiplicity} (blue). 
This case of $c = \Lambda_\text{had}/(2 m_0) = 1$ represents the largest multiplicity possible in the jet-like case, but even for large $M\sim100m_0$, the sizeable hierarchy between $m_0$ and $\Lambda_\text{had}$ results in only a handfull of produced glueballs per event. 
 This differs greatly from the high multiplicity production of pions in high-energy QCD jets.
 The blue band shows the range of multiplicities produced in simulated events, which is much more sharply peaked than a Poisson distribution. 
 We show the $N_c = 3$ case but the results are similar for other numbers of colours.

As a consistency check, we compare the average multiplicity predicted by \texttt{GlueShower} to the analytical expectation for average hadron multiplicity scaling in perturbative QCD. The standard result, for example found in \cite{Ellis:1991qj} for $N_f = 0$, is 
\begin{equation}
\langle N(E_{CM}^2) \rangle \propto \text{exp} \Bigg[ \frac{12\pi}{11C_A}\sqrt{\frac{2 C_A}{\pi \alpha(E_{CM}^2)}} + \frac{1}{4}\text{ln}\bigg(\alpha(E_{CM}^2)\bigg) \Bigg]~.
\label{mult eqn}
\end{equation}
This is normalized to the absolute multiplicity of our simulation for $M = 100 m_0$ and shown as the red dashed line in Fig.~\ref{fig:analytical multiplicity}. 
Note the good agreement for $M \gg m_0$,\footnote{Note that some deviation is expected since our shower uses 3-loop running of the coupling} but as expected, the scaling relation breaks down for $M$ closer to the glueball mass, demonstrating that finite-mass effects make analytical approximations of glueball distributions unreliable even for modest initial energies.

\begin{figure*}[t]
     \centering
     \begin{subfigure}{0.5\textwidth}
         \centering
         \includegraphics[width=\textwidth]{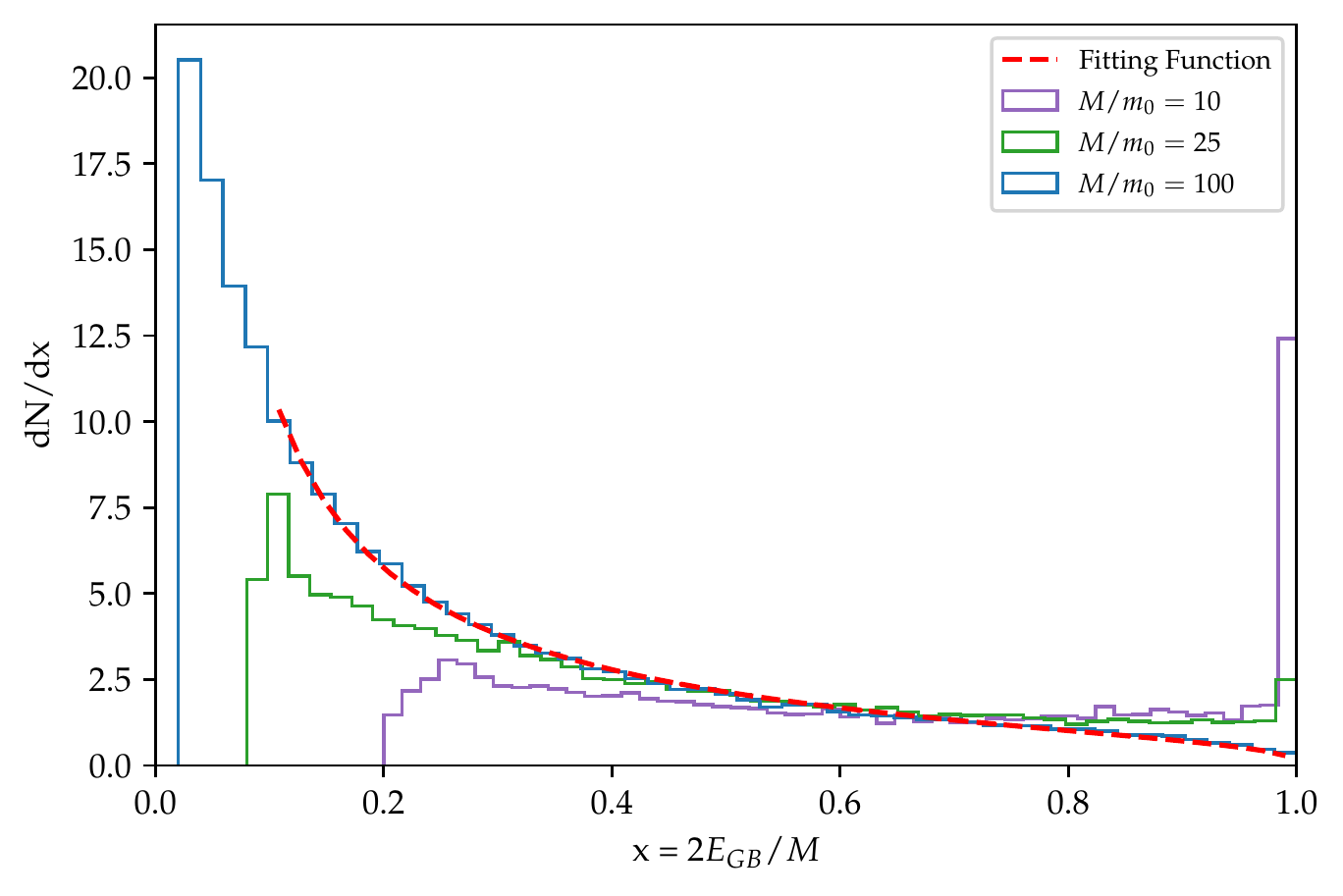}
         \label{fig:jet spectrum}
     \end{subfigure}%
     \begin{subfigure}{0.5\textwidth}
         \centering
         \includegraphics[width=\textwidth]{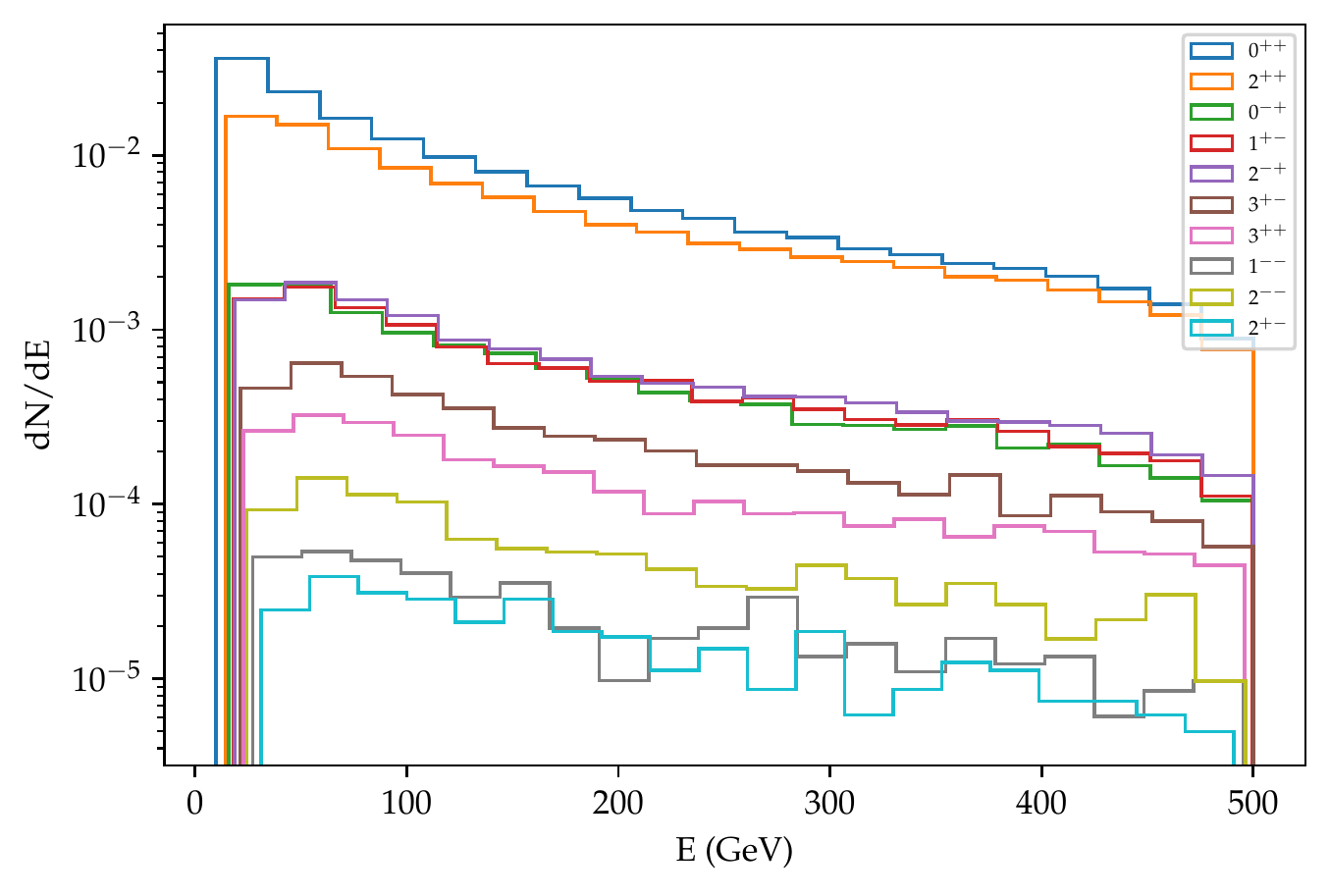}
         \label{fig:plasma spectrum}
     \end{subfigure}
        \caption{Glueball Energy Spectra for an example jet-like shower, with \texttt{plasma\_mode = False}, $\Lambda_\text{had}/(2m_0)=1$, $T_\text{had}/T_c=1$. \emph{Left:} Representative $0^{++}$ spectra for a range of $M$ values. The $M/m_0=100$ spectrum is fit to Eq.~\ref{eqn:highxfit} (red) on the range $0.1<x<1$.  \emph{Right:} Spectra for the lightest 10 glueball states for initial center-of-mass energy $M/m_0=100$.
        }
        \label{fig:all glueball spectrum}
\end{figure*}

\begin{figure*}[t]
     \centering
     \begin{subfigure}{0.5\textwidth}
         \centering
         \includegraphics[width=\textwidth]{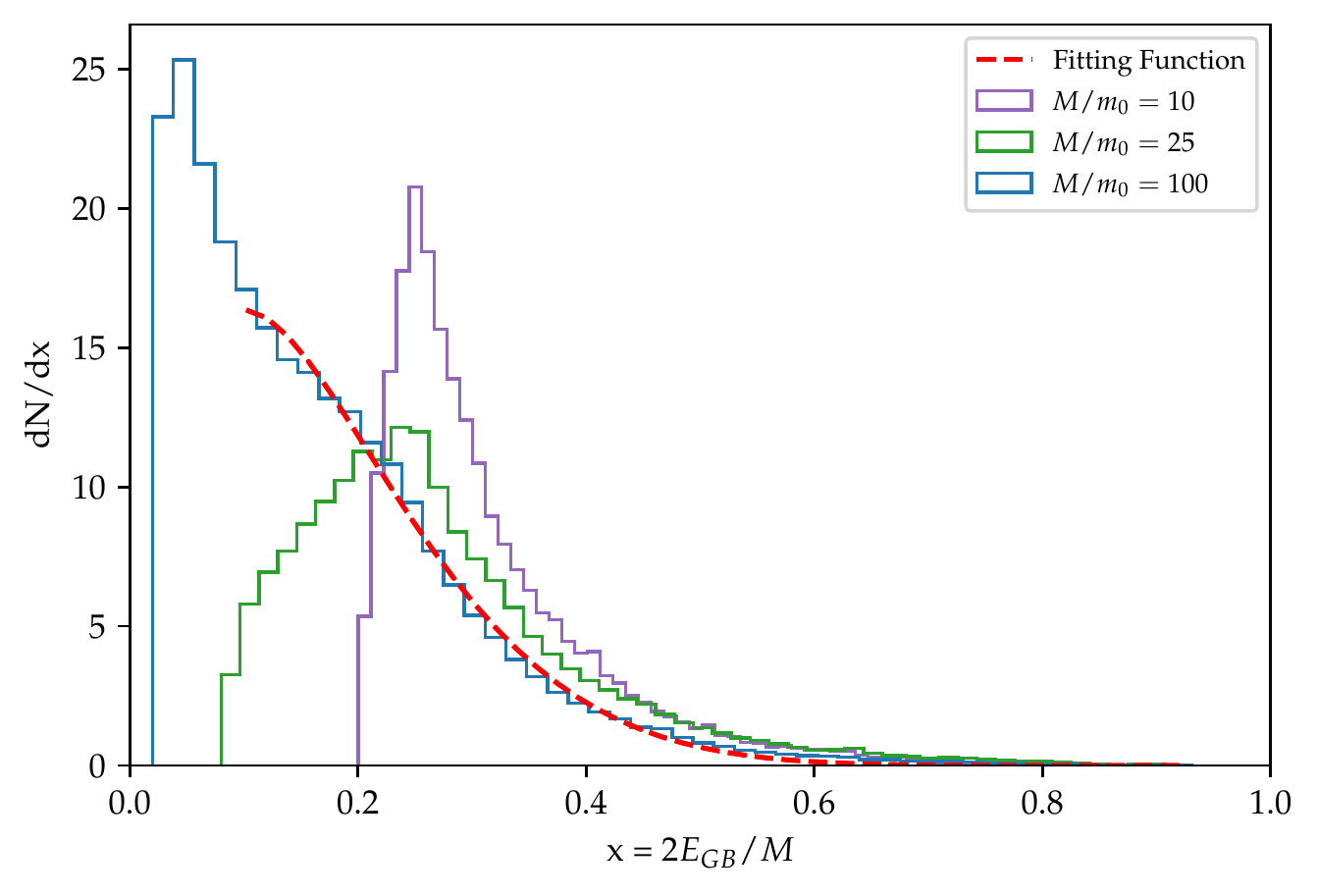}
         \label{fig:jet spectrum}
     \end{subfigure}%
     \begin{subfigure}{0.5\textwidth}
         \centering
         \includegraphics[width=\textwidth]{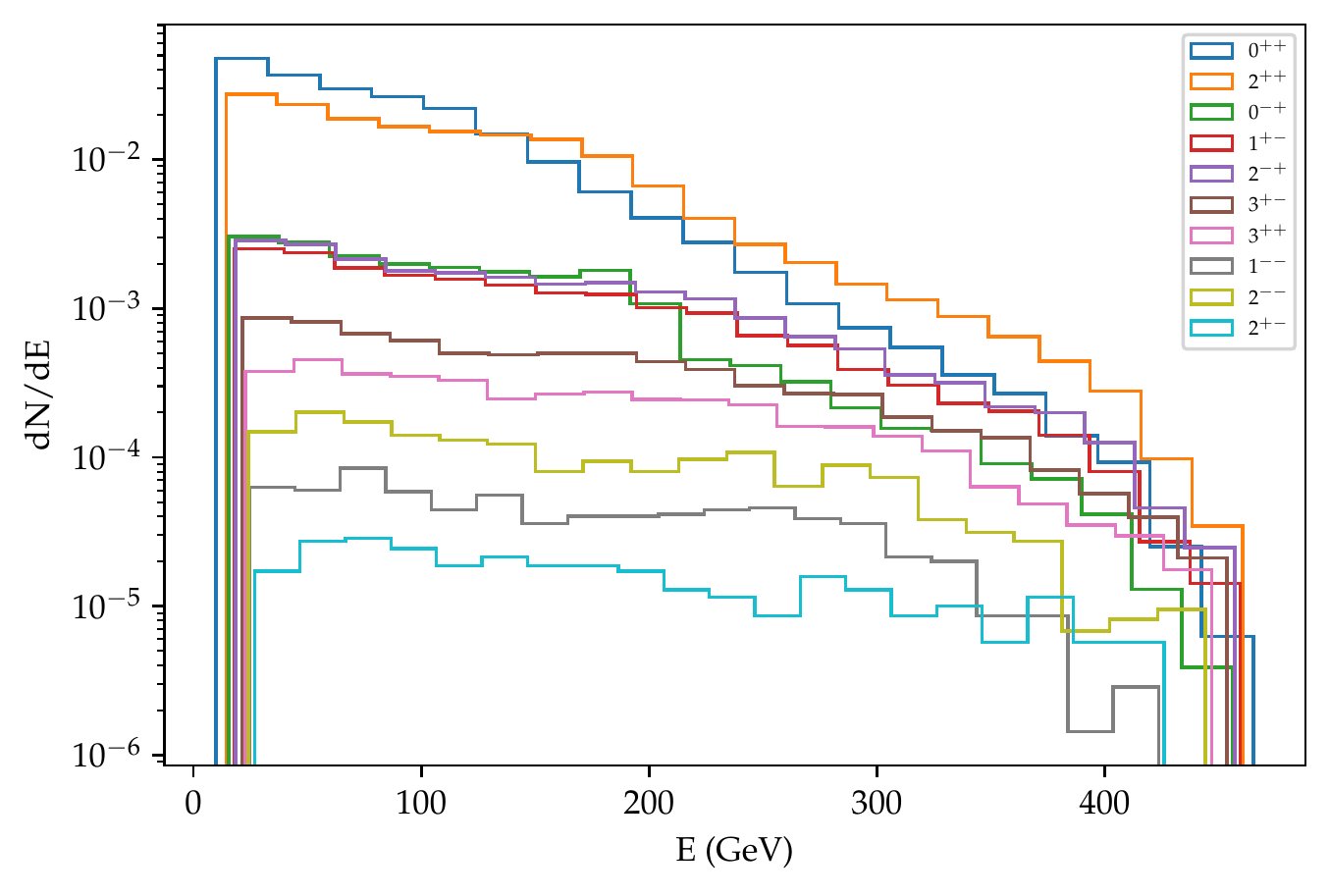}
         \label{fig:plasma spectrum}
     \end{subfigure}
        \caption{Glueball Energy Spectra for an example plasma-like shower, with \texttt{plasma\_mode = True}, $\Lambda_\text{had}/(2m_0)=4$, $T_\text{had}/T_c=1$. \emph{Left:} Representative $0^{++}$ spectra for a range of $M$ values. The $M/m_0=100$ spectrum is fit to Eq.~\ref{eqn:highxfit} (red) on the range $0.1<x<1$.  \emph{Right:} Spectra for the lightest 10 glueball states for initial center-of-mass energy $M/m_0=100$.
        }
        \label{fig:all glueball spectrum plasma}
\end{figure*}

We now compare the simulated glueball energy spectra to analytical expectations. 
In the SM, fragmentation functions are a priori unknown, thus a physically motivated functional form is used as an ansatz and fit to data.
We compare our jet-like output to one of the commonly used Colangelo and Nason function \cite{Colangelo:1992kh}
\begin{equation}
D_g^{GB}(x) \propto x^\alpha(1 - x)^\beta~,
\label{eqn:highxfit}
\end{equation}
where $x = 2E_{GB}/M$ and $E_{GB}$ is the glueball energy, thus $x\in [0,1]$. This function has been used for heavy quark fragmentation \cite{ParticleDataGroup:2020ssz} and is equivalent to the gluon-to-kaon fragmentation function parameterisation in \cite{deFlorian:2017lwf}, which has also been studied using lattice QCD~\cite{Salas-Chavira:2021wui}. 
Using a heavy quark fragmentation function for a pure glue shower might seem surprising, but this simply encodes that the energies of heavy hadronic final states are dominated by the heavy quark energies. This process is determined by the perturbative shower since it occurs significantly above the confinement scale, much like our treatment of jet-like glueball hadronization,  making such a functional form a reasonable ansatz. A similar approach was also used to analytically estimate dark glueball momenta in~\cite{Lichtenstein:2018kno}.

We compare our simulated events to an analytical fragmentation function in the high energy regime $M/m_0 = 100$ and for $0.1 < x < 1$~\cite{deFlorian:2017lwf}, where finite mass effects are less significant. As shown in Fig.~\ref{fig:all glueball spectrum} (left), we find very good agreement in this regime of applicability, provided we can find the required values of $\alpha$ and $\beta$ from data (i.e. simulation). 
In the same figure we show fragmentation functions for lower $M$, which demonstrates that the produced glueball spectrum becomes dominated by finite-mass effects as $M \to 10 m_0$, since small $x$-values become kinematically forbidden and a significant fraction of events only produce two glueballs with $x = 1$. 
This again demonstrates the limitations of using analytical approximations to estimate glueball distributions, and underlines the importance of using a self-consistent Monte Carlo simulation like \texttt{GlueShower}.

We also show the exclusive fragmentation functions of the lightest 10 glueball stats for $M/m_0 = 100$  in Fig.~\ref{fig:all glueball spectrum} (right).
For $T_\text{had} = T_c$, the final states predominately consist of $0^{++}$ and $2^{++}$ glueballs, followed by the next three heaviest states in roughly equal proportion. Heavier state production is suppressed by an order of magnitude.

\subsection{Plasma-like Hadronization}
We now discuss some of the same \texttt{GlueShower} outputs for a basic plasma-like shower case. The orange curve in Fig.~\ref{fig:analytical multiplicity} shows how multiplicity scales with initial center-of-mass energy for $\Lambda_\text{had}/(2 m_0) = c = 4$. 
Overall we find that the plasma-like case leads to higher multiplicity events compared to the jet-like case, but also that the inclusive multiplicity scales differently in various energy regimes:
\begin{itemize}
\item In the large energy limit $M \gg \Lambda_\text{had}$, we recover the same multiplicity scaling as the jet-like case, but larger by a constant since each produced plasma-ball evaporates into at least two glueballs. This does not mean, however, that the plasma-like case converges to the jet-like case in the high-energy limit: while the overall multiplicity scales similarly with energy, the absolute multiplicity is higher in the plasma-like case, resulting in lower characteristic energies for the final-state glueballs.\footnote{It is helpful to consider a SM QCD analogy: one could imagine a hypothetical hadronization model that dominantly produces heavy $B$-mesons (analogous to the plasma-balls), which then decay to lighter hadrons. Certainly, various multiplicity and momentum scalings of this $B$-meson-shower would be similar to the scalings of pion multiplicity and momenta in realistic SM hadronization, but the different hadronization assumption would also introduce fudamental differences in the final states no matter what the initial energy.}

\item For $M \to \Lambda_\text{had}$ the multiplicity asymptotes to a value larger than 2, in this case roughly 5. This is the regime in which the initial state only splits into two singlet plasma balls, which then each evaporate to at least two glueballs each.
\item At $M=\Lambda_\text{had}$ there is a discontinuity as the shower enters a new regime in which there is insufficient center-of-mass energy to form two separate plasma balls. For $M \leq \Lambda_\text{had}$, the initial di-gluon production event is therefore taken to form just a single colour-singlet plasma ball of mass $M$, which evaporates by glueball emission. Since this leads to a larger $m_\text{plasma}$ just below $\Lambda_\text{had}$ than just above, there is a small spike in produced glueball multiplicity, but this is a reasonable threshold effect. In this regime, glueball production is entirely SUEP-like, leading to a very different multiplicity scaling until the absolute minimum of $N=2$ is reached for $M < 3 m_0$. 
\end{itemize}


Inclusive and exclusive glueball fragmentation functions in the plasma-like case are shown in
Fig.~\ref{fig:all glueball spectrum plasma}.
As demonstrated in the left plot, the analytical functional form of the inclusive fragmentation function is still a fair approximation in the high-energy large-$x$ regime, though significantly worse than for the jet-like case, and for modest or low energies, the differences are even more pronounced.
On the right we show exclusive energy spectra for the lightest 10 glueball species, which are produced with very similar relative multiplicities as in the jet-like case. 
This is what we would expect, as relative glueball multiplicities are determined by the hadronization temperature which is kept constant between the cases considered in both Fig.~\ref{fig:all glueball spectrum} and Fig.~\ref{fig:all glueball spectrum plasma}. Additionally this shows that threshold effects in the jet-like case that favour the production of light glueballs have minimal effect.

\subsection{Defining Hadronization Benchmarks}

\begin{figure}[t]
\includegraphics[width=\linewidth]{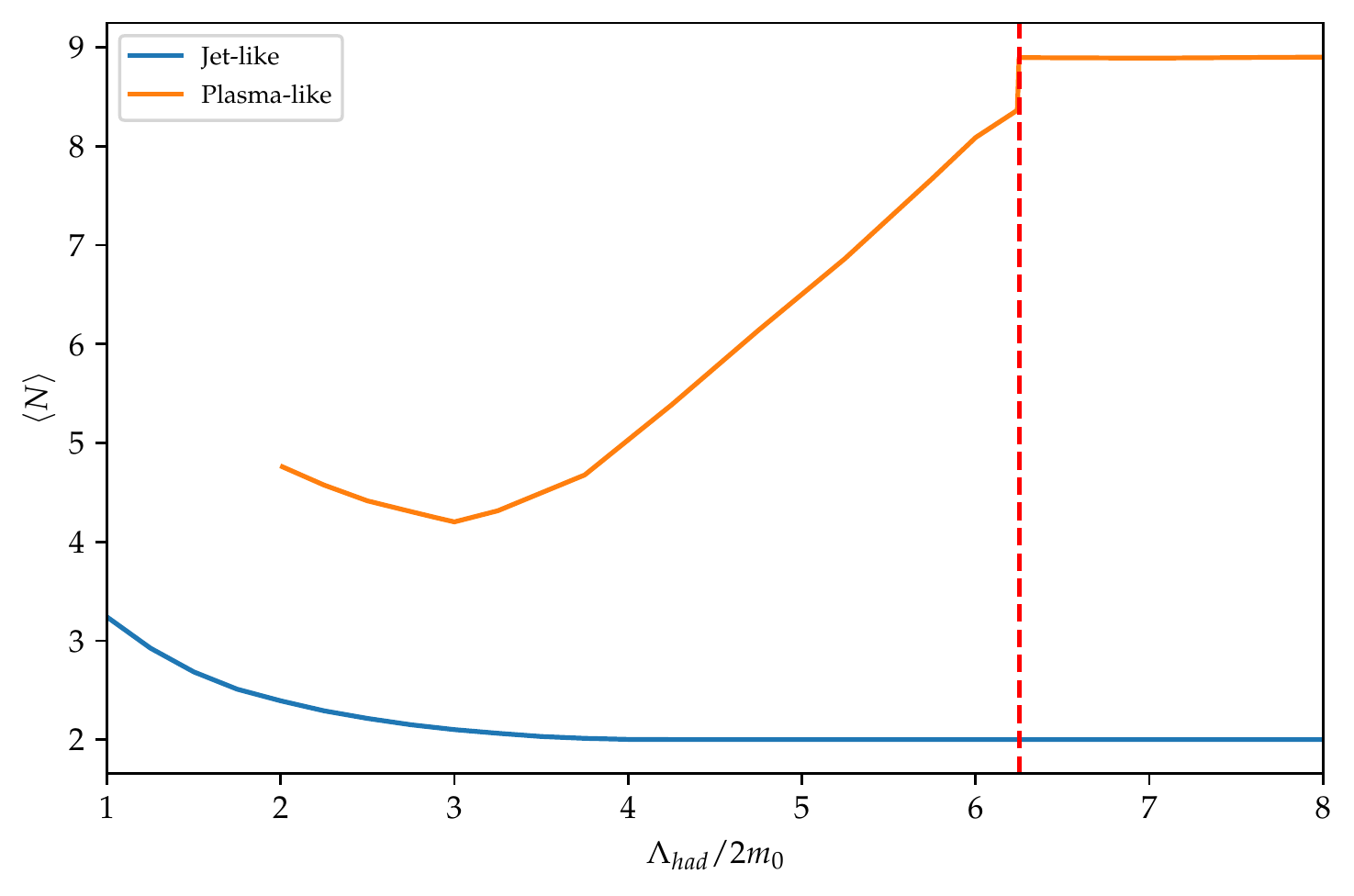}
\caption{Effect of $c = \Lambda_\text{had}/(2m_0)$ on average glueball multiplicity for exotic Higgs decays ($M=125 \gev$) into glueballs with $m_0 = 10 \gev$. 
In the jet-like case, $\Lambda_\text{had}$ is the hadronization scale at which the shower is terminated and gluons turned into glueballs. In the plasma-like case, it is the twice the mass of colour-singlet plasma balls produced in the shower, which then evaporate into glueballs. In both cases we take $T_\text{had}/T_c$=1. The red dashed line indicates $\Lambda_\text{had} = M$, resulting in the production of just two glueballs in the jet-like case and a single plasma ball of mass $M$ in the plasma-like case. 
}
\label{fig:c_range}
\end{figure}

Having discussed how some simple observables behave at different energies in both the jet-like and plasma-like cases, we now systematically examine their dependence on two nuisance parameters of our simulation: the hadronization scale $\Lambda_\text{had} = c \cdot 2m_0$ and the hadronization temperature $T_\text{had} = d  \cdot T_c$. This allows us to argue for a small set of benchmark hadronization parameters that should span the range of physically reasonable possible outcomes for glueball production, and which hence define theory uncertainties for physical predictions.

\begin{figure*}[t]
\includegraphics[width=\textwidth]{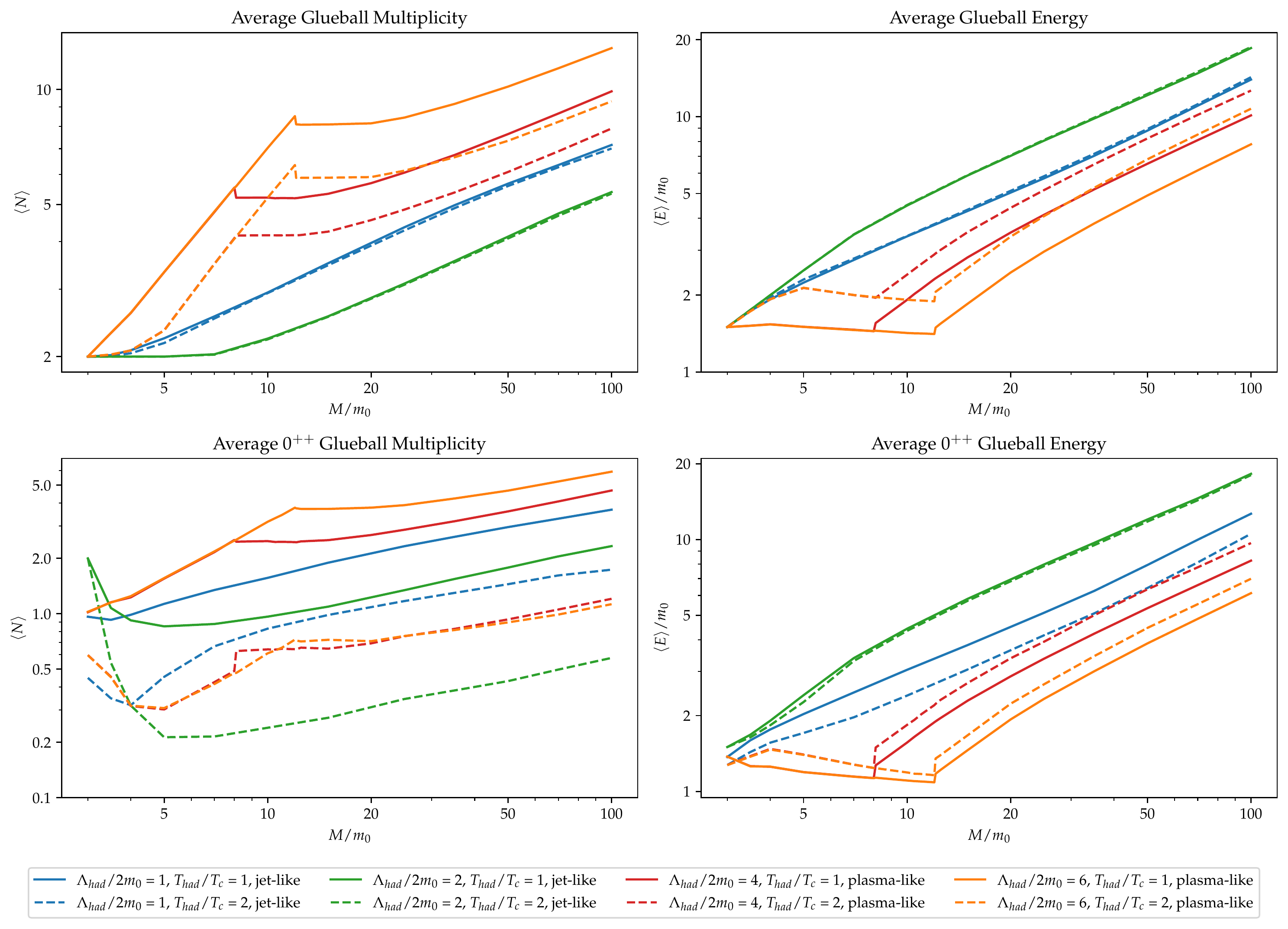}
\caption{Dependence of average glueball multiplicity and energy on $M/m_0$ for the 8 hadronization benchmarks with $N_c$ = 3. The range of predictions can be interpreted as our current theoretical uncertainty on glueball production: a factor of a few on average glueball energy and inclusive multiplicity, and a factor of 10 on exclusive  $0^{++}$ multiplicity.
}
\label{fig:four trend graphs}
\end{figure*}

\begin{figure*}[t]
\includegraphics[width=\textwidth]{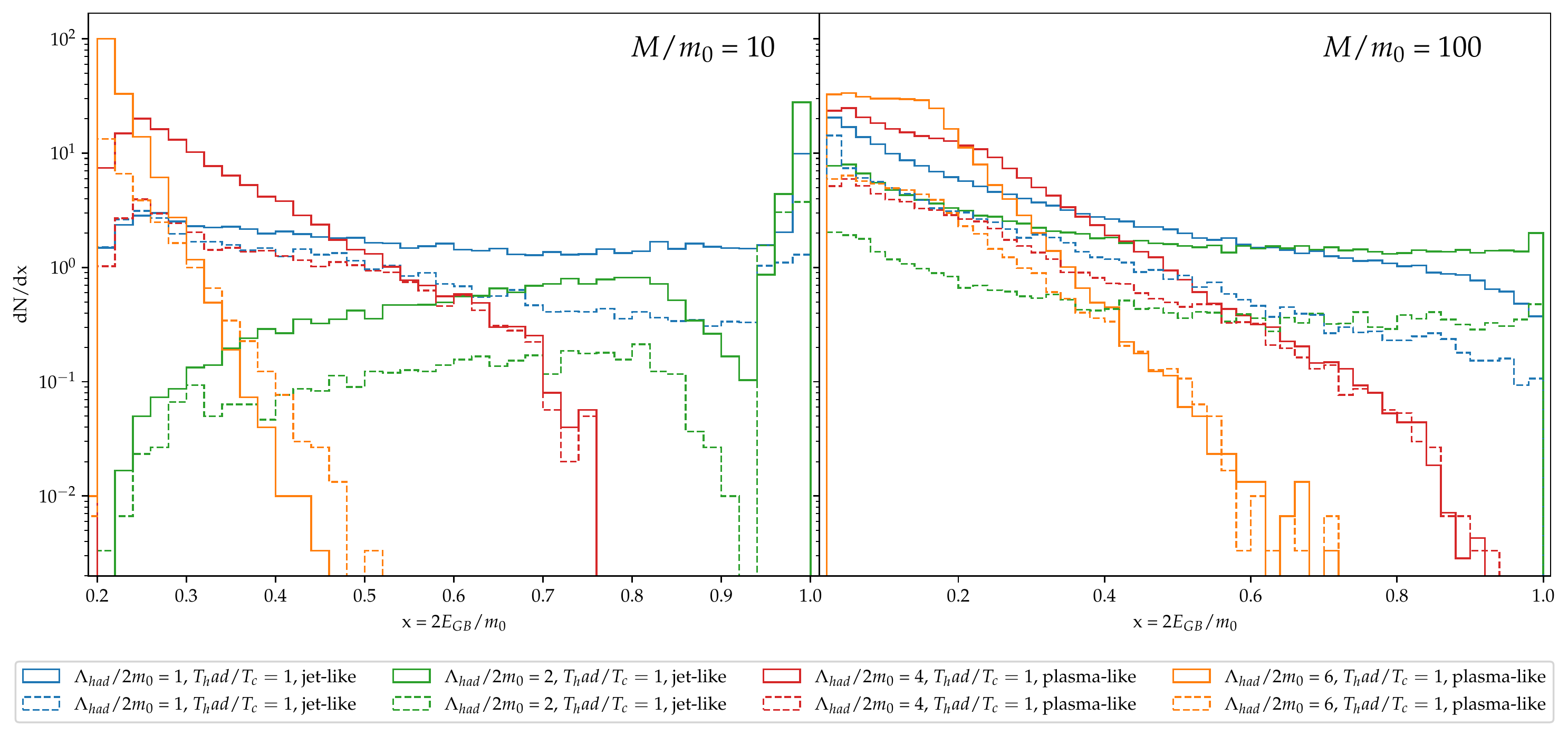}
\caption{$0^{++}$ glueball energy spectra for $M/m_0 = 10$ (left) and 100 (right) and $N_c = 3$, for our 8 hadronization benchmarks.}
\label{fig:frag func compare}
\end{figure*}

Figure~\ref{fig:c_range} shows how the inclusive glueball multiplicity in both the jet- and plasma-like cases depends on $\Lambda_\text{had}$ for $T_\text{had} = T_c$. 
In the jet-like case, shown in blue, we see that as $\Lambda_\text{had}$ increases the multiplicity decreases and asymptotes to 2. 
The plasma-like case (shown in orange) requires a minimum value of $c=2$ to allow each plasma ball to emit at least two glueballs. 
At this limit, only the lightest glueball can be produced. Naively we would expect multiplicity to increase with $c$, because even as increasing $c$ logarithmically suppresses the number of splittings in the shower, it linearly increases the number of glueballs emitted by the larger plasma states. Somewhat surprisingly, from $c=$ 2 to $\sim$ 3 the multiplicity decreases. This is due to the fact the plasma still only decays to two glueballs most of the time across this range, but now with greater access to the higher mass glueball states. Not until $c\sim4$ is the entire glueball spectrum sampled and produced in plasma ball decays, and further increasing $c$ now does lead to an increase in glueballs produced by the plasma.
Finally, as discussed in the previous subsection, when $\Lambda_\text{had} \to M$ only a single plasma ball is produced in the event, resulting in fully SUEP-like behaviour. 

While Fig.~\ref{fig:c_range} depicts both hadronization interpretations across the full range of $c$ values, each case is physically sensible in separate regimes. The jet-like interpretation assumes the initial flux tube fragments directly into final state glueballs, which is only sensible for values of $c$ close to 1. Larger values force gluons with virtualities $\sqrt{t} \gg m_0 \sim 6 \Lambda$  into a single low-mass glueball, which seems very implausible given the a priori reliable perturbative prediction for a much higher jet mass of that branch of the shower. We therefore adopt $c = 1, 2$ as benchmark values for the hadronization scale in the jet-like regime.

In contrast, the plasma-like case assumes fragmentation into high mass plasma states, and small values $c \lesssim 4$ seem to imply that these plasma balls have a large bias of evaporating only into the light glueball states. Sampling the full range of plasma-like behaviour therefore motivates picking slightly larger values of $c$. 
On the other hand, 
the plasma-like scenario already defies the most reasonable physical expectation based on our understanding of perturbative QCD and flux-tube dynamics, and a very large value of the plasma ball mass pushes this scenario into even more implausible regimes. For example, $c = 6$ corresponds to ending the perturbative shower at $\sim 70 \Lambda$, where $SU(N_c)$ should be entirely perturbative. It should therefore serve as a suitable ceiling for the possible range of plasma masses produced under this exotic assumption for the non-perturbative behaviour of $N_f = 0$ QCD. We therefore adopt $c = 4, 6$ as benchmark values for the hadronization scale in the plasma-like regime. 

The hadronization temperature $T_\text{had} = d \cdot T_c$ is less constrained by physical arguments and relatively unimportant for inclusive observables across glueball species. 
However, $T_\text{had}$ dominantly determines the exclusive predictions for each type of glueball. Given the exponential dependence of both glueball momentum in the plasma-like case and relative multiplicities in both jet- and plasma-like cases on $T_\text{had}$, sampling $d = 1, 2$ should span a wide range of physically plausible predictions.

In summary, careful phenomenological studies involving dark glueball production should compute physical predictions for a variety of different values of the nuisance parameters in \texttt{GlueShower}. For the jet-like case, $c = 1,2$ and $d = 1,2$ should be simulated. For the plasma-like case, $c = 4,6$ and $d = 1,2$. To be conservative, all 8 benchmark points should be used to define the systematic error bar on predictions.

\subsection{New predictions for Glueball Production}
We now have in hand a physically motivated simulation of the glueball production process, 
as well as 8 hadronization benchmarks which span the range of physically reasonable possible outcomes.
This allows us to make fully self-consistent predictions for glueball production with accurate theoretical uncertainties included for the first time. 

Naively, we would expect collider signals of dark glueball production to be most sensitive to the multiplicity and decay mode of the shortest-lived or most visibly-decaying glueball state, for example the $0^{++}$ if decay proceeds via the Higgs portal. 
Conversely,  indirect detection of dark matter annihilating into dark glueballs~\cite{Gemmell} will be affected by the relative distributions and decays of all unstable glueball species.
To give a feeling for how each of these two types of studies might be affected by theoretical uncertainties, we show in Fig.~\ref{fig:four trend graphs} how average multiplicity and energy  predictions for all glueballs inclusively and for the $0^{++}$ exclusively change across our range of possible hadronization benchmarks. We take the variation across all benchmarks to indicate the theoretical uncertainty for each observable. 

We find that average glueball multiplicity, average glueball energy and average $0^{++}$ energy have an uncertainty of about a factor of 3 across the range of considered center-of-mass energies $M/m_0$. 
On the other hand, the exclusive multiplicity of the $0^{++}$ state has a much larger spread of possible predictions, roughly a factor of 10 across the hadronization benchmarks. 
Slightly more can be said if one is willing to ascribe different priors to the default jet-like versus the more exotic plasma-like hadronization hypotheses.
Jet-like showers produce lower multiplicity jets of higher energy glueballs, while plasma-like showers produce fatter jets with higher multiplicities of softer glueballs. 

 Figure~\ref{fig:frag func compare} compares the $0^{++}$ energy spectra for different hadronization benchmarks. 
At low center-of-mass energies, we find very significant differences between the jet-like and plasma-like showers. While the plasma-like case favours low energy glueballs, the jet-like case is roughly flat, with dominant contributions by simple two-body glueball production. 
At high center-of-mass energies, the spectral shapes become much more similar, with low energy final states favoured, to varying degrees, across the benchmarks.

\section{Conclusion}
\label{conclusion}

Hidden Valleys are an extremely popular hypothesis for BSM physics. They may address fundamental mysteries like the hierarchy problem or the nature of dark matter, and their signatures are targeted by many new LHC searches (e.g.~\cite{Alimena:2019zri,ATLAS:2013bsk,ATLAS:2019tkk,CMS:2018bvr,CMS:2021dzg}) and proposed detectors \cite{Curtin:2018mvb,MATHUSLA:2020uve, Feng:2017uoz,Gligorov:2017nwh}.
However, the seemingly simple and minimal case of  $N_f = 0$ hidden QCD has undergone very little quantitative study, despite its high theoretical motivation within frameworks like 
Neutral Naturalness~\cite{Chacko:2005pe,Craig:2015pha,Burdman:2006tz,Barbieri:2005ri,Chacko:2005vw,Cai:2008au,Poland:2008ev,Cohen:2018mgv,Cheng:2018gvu}.
This can be traced back to our lack of understanding of pure glue hadronization. 

In this work, 
we show that significant progress can be made by combining a perturbative pure glue parton shower with a self-consistent and physically motivated parameterization of the unknown non-perturbative physics. This is in large part enabled by   the modest hierarchy between the glueball mass and the confinement scale $m_0/\Lambda \sim 6$ in $SU(N_c)$ theories. 
We make our simulation code available as the public Python code \texttt{GlueShower}, 
the first glueball generator for Hidden Valley theories.\footnote{\texttt{GlueShower} download: \href{https://github.com/davidrcurtin/GlueShower}{github.com/davidrcurtin/GlueShower}}
This can be combined with other Monte Carlo generators for dark gluon production and dark glueball decay~\cite{Juknevich:2009gg, Juknevich:2009ji} for a complete simulation of glueball production within a given dark sector scenario. 

Our approach incorporates what we consider to be an almost maximally wide range of possible hadronization scenarios, to make sure no physically reasonable possibility escapes our study. We argue that glueball production proceeds largely similar to jet-like hadron production in SM QCD, but we also consider the possibility that highly exotic non-perturbative physics of pure Yang-Mills theories somehow favours the production of high-mass colour-singlet gluon-plasma states, which evaporate via isotropic emission of thermal glueballs in their restframe. In both jet-like and plasma-like modes, a range of colour-singlet formation scales and hadronization temperatures can be selected to generate a range of possible hadronization behaviours. 
We intend \texttt{GlueShower} to be a starting point for $N_f = 0$ dark sector studies, and point out future improvements that could be implemented. A careful treatment of colour confinement or a study of gluons in the rope-like picture~\cite{Biro:1984cf} could lead to a more sophisticated hadronization model, but the current version represents a physically reasonable starting point for phenomenological studies.

We define a set of 8 hadronization benchmark scenarios, 4 in each of the above two modes, which we argue span the range of physically remotely reasonable hadronization outcomes for glueball production. 
We then use \texttt{GlueShower} to make some preliminary studies of glueball production observables, with theoretical uncertainties obtained from the variation across the different hadronization benchmarks. 
Uncertainties range from  modest $\mathcal{O}(1)$ factors to orders of magnitude, depending on the observable.

Accurately taking these uncertainties into account will be important for designing and interpreting future searches and constraints. In particular, our work could highlight which observables are more robust with respect to unknown details of glueball hadronization, encouraging a variety of collider and astrophysical probes to assemble a complete picture of the dark sector fundamental parameters and nonperturbative dynamics.  It is our hope that this work enables many  new studies and searches of dark sectors that were previously intractable. 

\vspace{5mm}

\emph{Acknowledgements:} 
We especially thank Matt Strassler for many insightful conversations and discussions. We also thank Jared Barron and Benjamin Fuks for helpful conversations. The research of DC and  CG was supported in part by a Discovery Grant from the Natural Sciences and Engineering Research Council of Canada, the Canada Research Chair program, the Alfred P. Sloan Foundation, and the Ontario Early Researcher Award. The work of CG was also supported by the University of Toronto Connaught International Scholarship.

\vspace{5mm}

\appendix 

\section{Perturbative QCD shower}
\label{app:pQCD}

We provide a step-by-step description of how \texttt{GlueShower} generates perturbative showers of gluons. 
This is a completely standard lowest-order perturbative shower with angular ordering, implemented following the Pythia manual~\cite{Sjostrand:2006za}, but we include this summary here for completeness and pedagogical purposes. 

The two initial gluons are treated separately to ensure energy-momentum conservation between their initial virtualities and energies. The shower is then evolved iteratively for all the subsequent daughter dark gluons.

\subsection{Initial Gluon Setup}
\begin{enumerate}
    \item Find virtuality of gluon 1 ($t_1$) assuming gluon 2 is on-shell. This is done by starting at the maximum allowed value, $(M - m_\text{min})^2$, and evolving down using the Monte Carlo method described in Sec.~\ref{perturbative shower}. There is some ambiguity regarding what on-shell means when the final state glueball species is still unspecified. For consistency across the shower, we define the minimum mass as half the hadronization scale, $c \cdot m_0$.
    \item Once $t_1$ is found, if this allows the gluon to split, $z_1$ is determined using the value of $t_1$.
    \item Steps 1 and 2 are repeated for gluon 2, to find $t_2$, and possibly $z_2$ if gluon 2 also splits.
    \item Using these values, a range of vetoes are checked before continuing with the shower.
    \begin{itemize}
        \item First check that $t_1+t_2<M$, if false then evolve the larger of the two virtualities to a smaller value.
        \item $z_{1,2}$ are found assuming that the other gluon was on-shell, but now they have virtualities of $t_{1,2}$; check that $z_{1,2}$ remain within the updated allowed range. If a gluon fails this check it virtuality is evolved to smaller values.
        \item Above vetoes are repeated until passed.
    \end{itemize}
    \item We now have values for the virtuality and splitting energy of each initial gluon, which are then used to evolve the daughters. Note that the 4-vectors of the initial gluons are now entirely determined.
\end{enumerate}

\subsection{Evolving the Daughters}
We now outline the steps applied for general gluon evolution at any point along the shower, excluding the initial gluons. We label the produced daughter gluons in this step 3 and 4.
\begin{enumerate}
    \item First, the coordinate system of this new splitting is established:
    \begin{itemize}
        \item The $z$ axis is aligned with the direction of the parent gluon's momentum.
        \item A random angle is chosen in the $x-y$ plane for the perpendicular momentum of the daughters.
    \end{itemize}
    \item Find initial $t$ and $z$ guess for the daughters.
    \begin{itemize}
        \item Initial energies are given by, $E_3 = z_1*E_1$ and $E_4 = (1 - z_1)*E_1$.
        \item The starting virtuality, $t_{\text{start},i}$~, for each daughter is given by min($E_i^2,m_1^2)$.
        \item Evolve the daughters down to find first guesses for $t_3$ and $t_4$. If splitting is allowed use these values to find $z_3$ and $z_4$.
    \end{itemize}
    \item
    Apply unconstrained/constrained evolution for each daughter. In unconstrained evolution, the maximum possible kinematic $z$ range is given when assuming the daughters have virtuality $m_\text{min}^2$. Thus, when the daughters are evolved and their actual virtualities are found, $t_{3,4}$, the originally assumed $z$ value may lie outside the newly determined range. The solution is to redefine the $z$ value. This is done by taking the original four momenta of the daughters and boosting them to their centre of mass frame, rescaling the vectors to reflect their determined evolved virtuality, and then boosting them back to the lab frame. This leads to a new $z$ value, matching the new energies of the rescaled dark gluons, that lies within the required range and is equivalent to using Eq.~(10.11) given in \cite{Sjostrand:2006za}. 
    
For constrained evolution, the final masses of the daughter gluons are constrained by the requirement that $z$ remain in the originally determined range. The default option enabled in the \texttt{GlueShower} code is unconstrained evolution, as used in Pythia, and leads to a higher rate of splittings in the shower.
    
    \item Impose angular ordering. This is the result of soft gluon coherence effects which cause the opening angle between the daughter gluons to be smaller than the opening angle of the parent gluon. This can be described as the supression of wide angle gluon emission. Practically it is enforced by determining the absolute maximum opening angle of the daughters, by assuming their daughters have mass $m_\text{min}$, and if this angle is bigger than the parent's opening angle, the daughter is evolved to  lower virtuality. Note that if a gluon is evolved to satisfy angular ordering, step 4 is repeated to ensure that the unconstrained evolution condition still holds.
    \item With the $t,z$ values of the daughter gluons determined, the steps of this section are repeated iteratively until all dark gluons in the shower have reached the hadronization scale.  
    \end{enumerate}

\bibliography{References}

\end{document}